\documentclass[a4paper,11pt]{article}
\usepackage{latexsym}
\usepackage{amsmath}
\usepackage{amssymb}
\usepackage{graphicx}
\usepackage{bm}
\usepackage{epsfig}
\usepackage{citesort}
\newcommand{\beq}{\begin{equation}}
\newcommand{\eeq}{\end{equation}}
\newcommand{\bea}{\begin{eqnarray}}
\newcommand{\eea}{\end{eqnarray}}

\begin{document}
\title{ Coupling of pion condensate, chiral condensate and
Polyakov loop in an extended NJL model }
\author{$\mbox{Zhao Zhang}^{b,a,1}$, $\mbox{Yu-xin
Liu}^{b,a,2}$\\[5pt]
\textit{${}^a$Department of Physics, Peking University, Beijing 100871, P. R. China}\\
\textit{${}^b$CCAST(World Laboratory), P.O. Box 8730, Beijing
100080, P. R. China} }
\date{\today}
\maketitle

\footnotetext[1]{E-mail: zhaozhang@pku.edu.cn}
\footnotetext[2]{E-mail: yxliu@pku.edu.cn}
\begin{abstract}

The Nambu Jona-Lasinio model with a Polyakov loop is extended to
finite isospin chemical potential case, which is characterized by
simultaneous coupling of pion condensate, chiral condensate and
Polyakov loop. The  pion condensate, chiral condensate and the
Polyakov loop as functions of temperature and isospin chemical
potential  are investigated  by minimizing the thermodynamic
potential of the system. The resulting $(T,\mu_I)$ phase diagram is
studied with emphasis on the critical point and Polyakov loop
dynamics. The tricritical point for the pion superfluidity phase
transition is confirmed and the phase transition for isospin
symmetry restoration in high isospin chemical potential region
perfectly coincides with the crossover phase transition for Polyakov
loop. These results are in agreement with the Lattice QCD data.

\vspace{10pt} PACS number(s): 12.38.Aw; 11.30.RD; 12.38.Lg;
\end{abstract}

\section{Introduction}
\vspace{5pt} Confinement and chiral symmetry breaking are the
fundamental properties of QCD in the nonperturbative domain. In
principle, the color liberation associated with  deconfinement is a
phenomenon distinguishable from the chiral symmetry restoration
phase transition,  but it is likely that the mechanism of
confinement would be closely related to chiral
dynamics\cite{Casher:1979}. It has been argued that the confined
vacuum necessarily breaks the chiral symmetry and the energy scale
of confinement is less then the energy scale of chiral symmetry
breaking: $\Lambda_{QCD}$ for confinement and
$\sim4\pi{f_\pi}$($f_\pi$ is the pion decay constant) for chiral
symmetry breaking\cite{Manohar:1984,Schaefer:1998}. In contrast with
the situation at zero temperature, the lattice data at finite
temperature suggest that the deconfinement phase transition and
chiral symmetry phase transition occur at the same
temperature\cite{Kogut:1983,Fukugita:1986,Karsch:1994,Aoki:1998,Karsch:2001,Gattringer:2002}.
Despite the attempts to explain above behaviors, the underlying
reason is still unknown.
\par
QCD phase diagram and thermodynamics has been the subject of intense
investigation in recent years. Lattice simulations is a principle
tool to explore the qualitative features of strongly interacting
matter and make quantitative prediction of its properties. Over the
years, this formulation has given us a wealth of information about
the phase diagram and thermodynamics at finite temperature and
limited chemical potential
\cite{Fodor:2003,Fodor:2002,Allton:2002,Allton:2003,Allton:2005,Laermann:2003,deForcrand:2003,delia1,delia2}.
In response to the lattice simulation, many phenomenological models
in terms of effective freedom degrees have been proposed to give an
interpretation of the available lattice data and to make prediction
in the region of phase diagram that can't be reached on the lattice
yet.
\par
As an effective chiral filed theory, classical Nambu-Jona-Lasinio
model (NJL)  \cite{Nambu,Klevansky:1992,Hatsuda:1994}  can
illustrate the transmutation of originally light quark or massless
quark into massive quansiparticles and generate the pions as
GoldStone bosons of spontaneously broken chiral symmetry at the same
time. This type models have also been used extensively to explore
color superconductivity phase transition at high baryon
density\cite{Buballa}. However, NJL models have a principle
deficiency. The reduction of the local color symmetry to the global
color symmetry leads to quarks not confined in standard  NJL. To
cure this problem, an improved field theoretical quasiparticle
model, a synthesis of Polyakov loop dynamics with NJL(PNJL), has
been proposed to interpret the Lattice QCD results and extrapolate
into the regions not accessible by lattice simulations
\cite{Meisinger:1996,Meisinger:2002,Fukushima:2004,Mocsy:2003,Megias:2004hj,Ratti1}.
For the chiral symmetry breaking, chiral condensate can be used as
an order parameter for chiral phase transition in the chiral limit.
In contrast, the deconfinement phase transition has a definite
meaning only at an infinite quark mass, which the expectation value
of Polyakov loop serves as an order parameter.  Since the $Z(3)$
center symmetry is explicitly broken in the real world with finite
quark mass,  no rigorous order parameter is established for
deconfinement phase transition\cite{Fukushima:2003}. However, the
Polyakov loop can still act as an indicator of a rapid crossover
towards deconfinement for finite quark mass. The motivation behind
the PNJL model is to unify the aspects of chiral symmetry breaking
and confinement through introducing  both the chiral condensate and
the Polyakov loop as classical, homogeneous fields which
simultaneously couple to quarks according the symmetry and symmetry
breaking patterns of QCD.
\par
The effectiveness of PNJL model has been tested in the literature by
confronting the PNJL results with the Lattice QCD data. It has been
reported that the two flavor PNJL model can reproduce the result
that the crossovers for deconfinement phase transition and the
chiral phase transition almost coincide\cite{Fukushima:2004,Ratti1}.
For finite quark chemical potential, further investigations also
suggest that the thermodynamics and susceptibilities obtained in
PNJL model are excellently  in agreement with the corresponding
Lattice QCD data\cite{Ratti1,Ghosh:2006,Ratti2}. In addition, the
PNJL model has been extended to include the diquark degrees of
freedom and used to explore the phase diagrams in high baryon
chemical potential\cite{Simon}. Though the PNJL results have a
satisfactory agreement with some lattice data, there still need to
perform further test of the PNJL since for finite baryon chemica
potential the available lattice simulation is limited only in the
low region.
\par
It is well known that there is no sigh problem for the lattice
calculation for the case with finite isospin chemical potential and
zero baryon chemical
potential\cite{Son:2001,Kogut:2001,Kogut:2002,Kogut:2004}.
Therefore, it is interesting to investigate the phase diagram and
thermodynamics in the framework of PNJL model at finite isospin
chemical potential and compare the results with the corresponding
lattice data. In \cite{Mukherjee}, the authors have extended the
PNJL to the case with low baryon and isospin chemical potential
without considering the pion condensate. Both the QCD effective
filed theory\cite{Son:2001} and lattice
calculations\cite{Kogut:2001,Kogut:2002,Kogut:2004} have shown that
 the pion superfluidity phase (charged pion Boson-Einstein condensate) may
occur in high enough isospin chemical potential. This phase had also
been explored by other chiral models of QCD such as Ladder
QCD\cite{LADD}, NJL model\cite{NJL1,NJL3,NJL7}, random matrix
model\cite{MAT2} and global color model\cite{GCM}. Therefore, it is
necessary to extend the PNJL model to high isospin density region to
investigate the influence of Polyakov loop dynamics  on the phase
structure by including the pion condensate degrees of freedom and
give further test of the validity of the PNJL by confronting the
corresponding Lattice QCD data. In addition, due to the limitation
of lattice simulation, the talk between effective fields theory and
Lattice QCD is needed.
\par
The primary purposes of this paper is to derive the formulation of
the PNJL model with simultaneously considering the pion condensate,
chiral condensate and Polyakov loop degrees of freedom at mean filed
level and then use it to explore the phase diagram of two flavor QCD
to compare with the corresponding lattice data. The emphases will be
put on the critical point and the impact of Polyakov loop dynamics.
For simplicity, the freedom degrees of diquark is not included in
this paper and the baryon chemical potential is limited in low
region.

\section{PNJL model at finite isospin chemical potential}

Extending the work \cite{Ratti1} to  finite baryon chemical
potential and isospin chemical potential case, the lagrangian of
two-flavor PNJL model is given by
\begin{eqnarray}
\mathcal{L}_{PNJL}=\bar{\psi}\left(i\gamma_{\mu}D^{\mu}+\gamma_0\hat{\mu}
-\hat{m}_0-i\hat{\lambda}\gamma_5\tau_1\right)\psi+G\left[\left(\bar{\psi}\psi\right)^2+\left(\bar{\psi}i\gamma_5
\vec{\tau}\psi \right)^2\right]
-\mathcal{U}\left(\Phi[A],\bar{\Phi}[A],T\right), \label{lagr}
\end{eqnarray}
where $\psi=\left(\psi_u,\psi_d\right)^T$ is the quark field,
\begin{eqnarray}
D^{\mu}=\partial^\mu-iA^\mu~~~~~~\mathrm{and}~~~~~~A^\mu=\delta_{\mu0}A^0~~.
\end{eqnarray}
The two-flavor current quark mass matrix is $\hat{m}_0 = diag(m_u,
m_d)$ and we shall work in the isospin symmetric limit with $m_u =
m_d \equiv m_0$. The quark chemical potential matrix $\hat{\mu}$
takes the form
\begin{equation}
\hat{\mu}=\bigg(\begin{array}{cc}
    \mu_u & \\
     & \mu_d\end{array}
 \bigg)=\bigg(\begin{array}{cc}
    \mu+\mu_I & \\
     & \mu-\mu_I\end{array}
 \bigg),
\end{equation}
with
\begin{eqnarray}
\mu=\frac{\mu_u+\mu_d}{2}=\frac{\mu_B}{3}~~~~~~\mathrm{and}~~~~~~\mu_I=\frac{\mu_u-\mu_d}{2},
\end{eqnarray}
where $\mu_B$ is baryon chemical potential corresponding to
conserved baryon charge and $\mu_I$ is isospin chemical potential
corresponding to conserved isospin charge.  A term proportional to
$\lambda$ is introduced in (\ref{lagr}) which explicit break the
$U_{I_3}(1)$ symmetry (Below we call it $I_3$ symmetry).  A local,
chirally symmetric scalar-pseudoscalar four-point interaction of the
quark fields is introduced with an effective coupling strength $G$.
\par
In comparison with the classical NJL lagrangian, the gauge field
term $A^\mu(x) = g {\cal A}^\mu_a(x)\frac{\lambda_a}{2}$ is
contained and an effective potential
$\mathcal{U}(\Phi,\bar{\Phi},T)$ expressed in terms of the traced
Polyakov loop $\Phi=(\mathrm{Tr}_c\,L)/N_c$ and its (charge)
conjugate $\bar{\Phi}=(\mathrm{Tr}_cL^{\dagger})/N_c$ in the PNJL
lagrangian. The Polyakov loop $L$ is a matrix in colour space
explicitly given by
\begin{equation}
L\left(\vec{x}\right)=\mathcal{P}\exp\left[i\int_{0}^{\beta}d\tau\,A_4\left(\vec{x},\tau\right)\right],
\end{equation}
with $\beta = 1/T$ the inverse temperature and $A_4 = i A^0$. In a
convenient gauge (the so-called Polyakov gauge), the Polyakov loop
matrix can be given a diagonal representation~\cite{Fukushima:2003}.
The coupling between Polyakov loop and quarks is uniquely determined
by the covariant derivative $D_\mu$ in the PNJL
Lagrangian~(\ref{lagr}). For simplicity,  the temporal component of
Euclidean gauge field $A_4$ is treated as a constant in PNJL, and
the Polyakov loop is reduced as
\begin{equation}
L=\left[\mathcal{P}\exp\left(i\int_{0}^{\beta}
A_4d\tau\right)\right]=\exp\left[\frac{iA_4}{T}\right] \label{l}.
\end{equation}
 Corresponding to above expression, the trace of the Polyakov
loop, $\Phi$, and its conjugate, $\bar{\Phi}$, are treated as
classical field variables in PNJL.
\par
The temperature dependent effective potential
$\mathcal{U}(\Phi,\bar{\Phi},T)$ is used to mimic pure-gauge Lattice
QCD data,  which should have exact $Z(3)$ center symmetry. In this
paper, we will use the potential $\mathcal{U}(\Phi,\bar{\Phi},T)$
proposed in \cite{Ratti1} , which takes the form \bea
{\mathcal{U}\left(\Phi,\bar{\Phi},T\right)\over T^4}
=-{b_2\left(T\right)\over 2}\bar{\Phi} \Phi- {b_3\over
6}\left(\Phi^3+ {\bar{\Phi}}^3\right)+ {b_4\over 4}\left(\bar{\Phi}
\Phi\right)^2 \label{u1} \eea with \beq
b_2\left(T\right)=a_0+a_1\left(\frac{T_0}{T}\right)+a_2\left(\frac{T_0}{T}
\right)^2+a_3\left(\frac{T_0}{T}\right)^3~~~. \label{u2} \eeq A
precision fit of the coefficients $a_i,~b_i$ is performed to
reproduce the lattice data and  $T_0=270\mathrm {MeV}$ is adopted
in this paper.
\par
To describe the spontaneous chiral symmetry breaking and $I_3$
symmetry breaking, we define the chiral condensate as
\begin{equation}
\langle{\bar{\psi}\psi\rangle}=\sigma=\sigma_u+\sigma_d,
\end{equation}
with $\sigma_u=\langle{\bar{u}u}\rangle$ and
$\sigma_d=\langle{\bar{d}d}\rangle$, and the charged pion
condensates
\begin{equation}
\langle{\bar{\psi}i\gamma_5\tau_{+}\psi}\rangle=\pi^{+}=\frac{\pi}{\sqrt{2}}e^{i\theta},
\quad
\langle{\bar{\psi}i\gamma_5\tau_{-}\psi}\rangle=\pi^{-}=\frac{\pi}{\sqrt{2}}e^{-i\theta},
\end{equation}
where $\tau_{\pm}=(\tau_1\pm\tau_2)/\sqrt{2}$ and $\tau_i$ is the
Pauli matrix in flavor space. In the chiral limit, nonzero
condensate $\sigma$ indicates spontaneous chiral symmetry breaking,
and nonzero condensate $\pi$ indicates spontaneous $I_3$ symmetry
breaking. The phase factor $\theta$ describes the direction of the
 $I_3$ symmetry breaking. For convenience, we adopt
$\theta=0$ in the following and the pion condensate can be expressed
as
\begin{equation}
\langle{\bar{\psi}i\gamma_5\tau_1\psi}\rangle=\pi.
\end{equation}
This choice of the $I_3$ symmetry breaking direction  is consistent
with the form of the explicit isospin breaking term introduced in
(\ref{lagr}).
\par
Using the standard bosonization techniques, the mean filed
lagrangian of PNJL takes the form
\begin{equation}
\mathcal{L}_{\mathrm{mf}}=\bar{\psi}\left(i\gamma_{\mu}D^{\mu}+\gamma_0\hat{\mu}
-M-Ni\gamma_5\tau_1\right)\psi-G\left[\sigma^2+\pi^2\right]
-\mathcal{U}\left(\Phi[A],\bar{\Phi}[A],T\right),
\end{equation}
with the gaps
\begin{eqnarray}
M=m_0-2G\sigma,\\
N=\lambda-2G\pi.
\end{eqnarray}
The thermal dynamical potential in the mean filed level is expressed
as
\begin{eqnarray}
\Omega&=&{\cal
U}\left(\Phi,\bar{\Phi},T\right)+G(\sigma^2+\pi^2)-\frac{T}{V}\mathrm{ln} \mathrm{det}S^{-1}_{\mathrm{mf}}\nonumber\\
&=&{\cal
U}\left(\Phi,\bar{\Phi},T\right)+G(\sigma^2+\pi^2)-T\sum_n\int\frac{d^3p}{\left(2\pi\right)^3}
\mathrm{Tr}\ln{\frac{{S}_{mf}^{-1}\left(i\omega_n,\vec{p}\,\right)}{T}}
,\nonumber\\
\label{omega}
\end{eqnarray}
where  the sum is taken over Matsubara frequencies
$w_n=(2n+1)\pi{T}$ and the trace is taken over color, flavor and
Dirac indices. The momentum dependent inverse quark propagator
matrix including both chiral and pion condensates in flavor space
takes the form
\begin{equation}
S^{-1}_{\mathrm{mf}}(iw_n,\vec{p})=\bigg(\begin{array}{cc}
    (iw_n+\mu+\mu_I+iA_4)\gamma_0-\vec{\gamma}\cdot\vec{p}-M&-i\gamma_5N\\
    -i\gamma_5N& (iw_n+\mu-\mu_I+iA_4)\gamma_0-\vec{\gamma}\cdot\vec{p}-M
    \end{array}\bigg).
\end{equation}
Using the identity $\mathrm{Tr\ ln(X)}=\mathrm{ln}\ \mathrm{det(X)}$
and the technique
\begin{equation}
\mathrm{det}\bigg(\begin{array}{cc}
    A & B\\
    C & D\end{array}
 \bigg)=\text{det}(A)\text{det}(B)\text{det}(C)\text{det}(C^{-1}DB^{-1}-A^{-1}),
\end{equation}
the thermal dynamical potential(\ref{omega}) can be further
expressed as
\begin{eqnarray}\label{thermalp}
\Omega&=&G(\sigma^2+\pi^2)+{\cal
U}(\Phi,\bar{\Phi},T)-2N_c\int{\frac{d^3p}{(2\pi)^3}}\big[E_p^-+E_p^+\big]\theta(\Lambda^2-\vec{p}^2)\nonumber\\
&&-2T\int{\frac{d^3p}{(2\pi)^3}}
\mathrm{Tr_c}\Big[\text{ln}[1+L^+e^{-(E_p^-+\mu)/T}]
+\text{ln}[1+Le^{-(E_p^--\mu)/T}]\nonumber\\
&&+\text{ln}[1+{L^+}e^{-({E_p^+}+\mu)/T}]
+\text{ln}[1+Le^{-(E_p^+-\mu)/T}]\Big],\label{thermp}
\end{eqnarray}
with quasiparticle energy  $E_p^{\pm}=\sqrt{(E_p\pm\mu_I)^2+N^2}$
and $E_p=\sqrt{\vec{p}^2+M^2}$. Performing the remaining color
trace, the integrand of last term on r.h.s. (\ref{thermp}) is given
by
\begin{eqnarray}\label{factorout}
\nonumber
&&\ln\mathrm{det}\left[1+L\,\mathrm{e}^{-\left(E_p^--\mu\right)/T}\right]+
\ln \mathrm{det}\left[1+L^{+}\,\mathrm{e}^{-\left(E_p^-+\mu\right)/T}\right]\\
\nonumber
&&+\ln\mathrm{det}\left[1+L\,\mathrm{e}^{-\left(E_p^+-\mu\right)/T}\right]+
\ln \mathrm{det}\left[1+L^{+}\,\mathrm{e}^{-\left(E_p^++\mu\right)/T}\right]\\
&=&\ln\left[1+3\left(\Phi+\bar{\Phi}\mathrm{e}^{-\left(E_p^--\mu\right)/T}\right)\mathrm{e}^{-\left(E_p^--\mu\right)/T}+
\mathrm{e}^{-3\left(E_p^--\mu\right)/T}\right]\nonumber\\
&+&\ln\left[1+3\left(\bar{\Phi}+\Phi\mathrm{e}^{-\left(E_p^-+\mu\right)/T}\right)\mathrm{e}^{-\left(E_p^-+\mu\right)/T}+
\mathrm{e}^{-3\left(E_p^-+\mu\right)/T}\right]\nonumber\\
&+&\ln\left[1+3\left(\Phi+\bar{\Phi}\mathrm{e}^{-\left(E_p^+-\mu\right)/T}\right)\mathrm{e}^{-\left(E_p^+-\mu\right)/T}+
\mathrm{e}^{-3\left(E_p^+-\mu\right)/T}\right]\nonumber\\
&+&\ln\left[1+3\left(\bar{\Phi}+\Phi\mathrm{e}^{-\left(E_p^++\mu\right)/T}\right)\mathrm{e}^{-\left(E_p^++\mu\right)/T}+
\mathrm{e}^{-3\left(E_p^++\mu\right)/T}\right].
\end{eqnarray}
From (\ref{thermalp}) and (\ref{factorout}), we see that the trace
of the Polyakov loop $\Phi$ and its conjugate $\bar{\Phi}$ can still
be factored out despite the appearance of the off-diagonal terms in
the inverse quark propagator. It is easy seen from (\ref{factorout})
that the classical NJL model is restored from PNJL when $\Phi$ and
$\bar{\Phi}$ approach $1$.
\par
Minimizing the thermal dynamical potential (\ref{thermalp}), the
motion equations for the mean fields $\sigma$, $\pi$, $\Phi$ and
$\bar{\Phi}$ are determined through the coupled equations
\begin{equation}
\frac{\partial\Omega}{\partial\sigma}=0,\quad\frac{\partial\Omega}{\partial\pi}=0,
\quad\frac{\partial\Omega}{\partial\Phi}=0,\quad\frac{\partial\Omega}{\partial\bar{\Phi}}=0.
\end{equation}
This set of equations is then solved for the fields $\sigma$, $\pi$,
$\Phi$ and $\bar{\Phi}$ as functions of  the temperature $T$, baryon
chemical potentials $\mu_B$ and isospin chemical potential $\mu_I$.
Note that when there exist multi roots of these coupled equations,
the solution corresponding to the minimal thermodynamical potential
is favored.
\par
The NJL part of the model involves three parameters: the current
quark mass of $u$ and $d$, the local four fermion coupling constant
$G$ and the three-momentum cutoff $\Lambda$. In this work, these
parameter are fixed to reproduce three physical quantities in
vacuum: pion mass $m_\pi=140\mathrm{MeV}$, pion decay constant
$f_\pi=93\mathrm{MeV}$ and chiral condensate
$\langle\bar{u}u\rangle=\langle\bar{d}d\rangle=\sigma_0/2=-(250\mathrm{MeV})^3$
with
\begin{equation}
m_0=5.5\mathrm{MeV},\quad G=5.04 \mathrm{GeV}^{-2},\quad
\Lambda=0.651 \mathrm{GeV}.
\end{equation}
\par
Once $\Omega$ is known, the thermodynamic functions that measure the
bulk properties of matter can be obtained. For an infinite system,
the pressure $p$, the entropy density $s$, the baryon number density
$n_B$, the isospin number density $n_I$, the flavor number density
$n_u$ and $n_d$, the energy density $\epsilon$ and the specific heat
$c$ are all derived from $\Omega$.

\par
For $\mu_B=0$, the trace of Polyakov loop $\Phi$ is equal to its
conjugate $\bar{\Phi}$ and both fields are real. Without concerning
the isospin chemical potential, numerical results suggests that the
crossover phase transitions for the chiral condensate $\sigma$ and
for the Polyakov loop $\Phi$ almost coincide in PNJL and the
difference between the two transition temperatures is within
$5\mathrm {MeV}$\cite{Ratti1,Ghosh:2006}. Following\cite{Ratti1}, we
define $T_c$ as the average of the two transition temperature and we
verified the result that $T_c^{0}=227\mathrm{MeV}$ \cite{Ghosh:2006}
for $\mu_B=\mu_I=0$. Since the PNJL model uses a simple static
background field representing the Polyakov loop and reduce the gluon
dynamics to chiral point couplings between quarks, this model can
not be expected to work beyond a limited temperature and quark
chemical potential. It is assumed this model works up to an upper
limit of temperature with $T\leq(2-3)T_c^{0}$ because the transverse
gluon degrees of freedom will be important for
$T>2.5T_c^{0}$\cite{Meisinger:2004}. In this paper we only consider
the case for $\mu_B=0$ and we assume that PNJL works well in the
region $|\mu_I|<500\text{MeV}$

\section{Results}
\vspace{5pt}
\subsection{For $\lambda=0$ and $\mu_B=0$}
For $\lambda=0$, the charged pion condensate $\pi$ is the true order
parameter for the pion superfluidity phase transition. Previous
investigations based on the QCD effective field
theory\cite{Son:2001} and lattice
simulations\cite{Kogut:2001,Kogut:2002,Kogut:2004} have shown that
pion superfluidity phase occurs when $\mu_I>\mu_c=m_\pi/2$ for $T=0$
and $\mu_B=0$, where the phase transition is second order.  At
sufficient high temperature and $\mu_I>\mu_c$, the pion condensate
evaporates and the spontaneous breaking $I_3$ symmetry and parity
are restored. The lattice calculations also suggest that the pion
condensate vanishes for sufficient high $\mu_I$ and this $I_3$
symmetry restoration phase transition may be first order for large
enough $\mu_I$ . Therefore, there should exist a tricritical point
(TCP) in the $(T,\mu_I)$ phase diagram. We will show below that
these conclusions can be verified in our calculations within the
PNJL model.

\begin{figure}[ht]
\hspace{-.05\textwidth}
\begin{minipage}[t]{.4\textwidth}
{\includegraphics*[width=\textwidth]{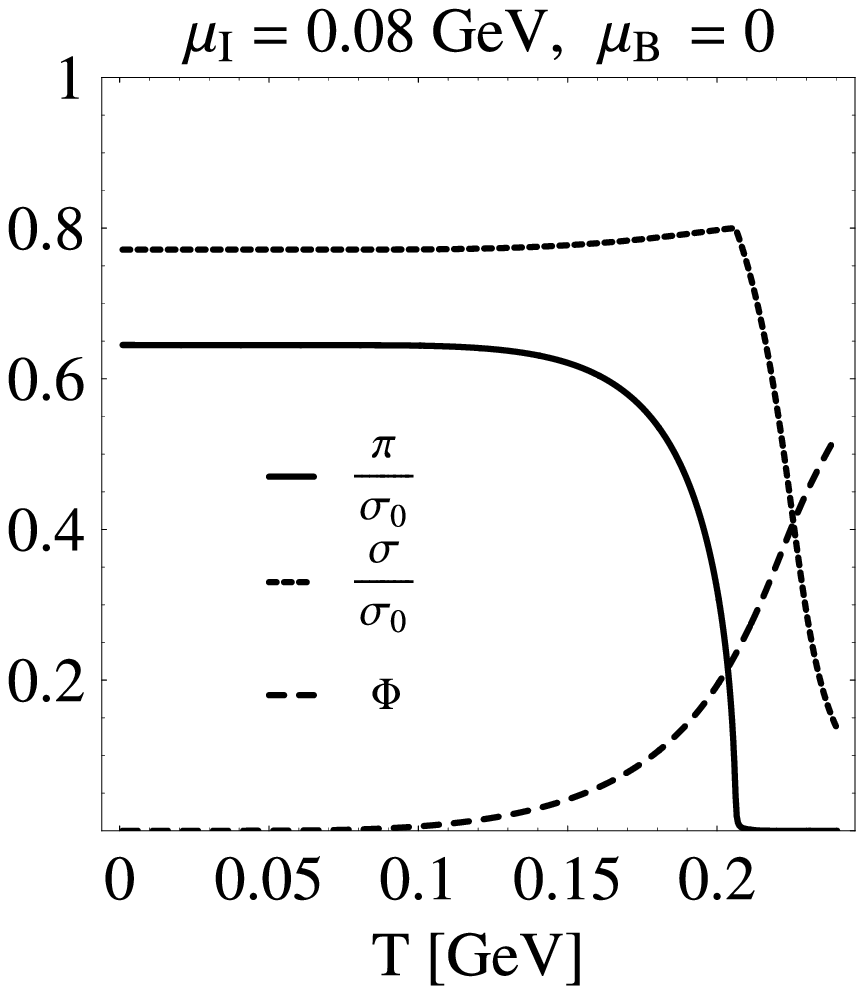}}\\
\centerline{(a)}
\end{minipage}
\hspace{.05\textwidth}
\begin{minipage}[t]{.4\textwidth}
\scalebox{.85}{
\includegraphics*[width=1.17\textwidth]{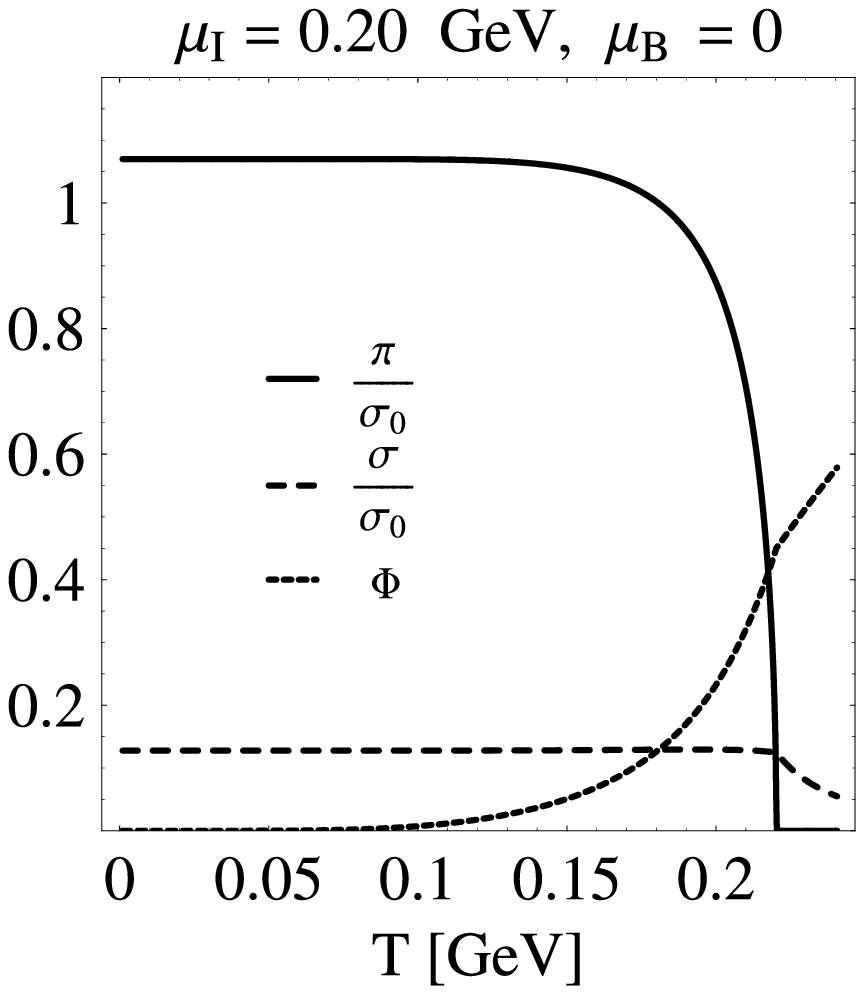}}\\
\centerline{(b)}
\end{minipage}
\centering
\begin{minipage}[t]{.4\textwidth}
\scalebox{.85}{
\includegraphics*[width=1.17\textwidth]{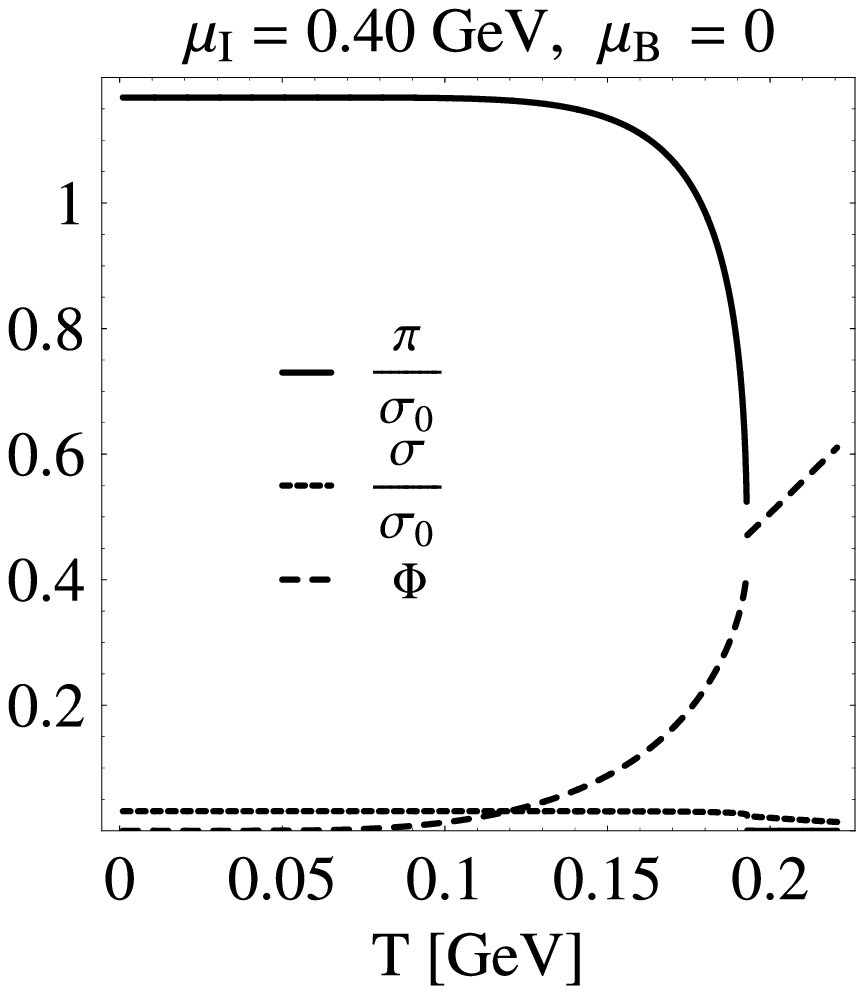}}\\
\centerline{(c)}
\end{minipage}
\parbox{15cm}
{\caption{Scaled pion condensate, chiral condensate and Polyakov
loop $\Phi$ as functions of temperature at zero baryon chemical
potential for different $\mu_I$ with  $\lambda=0$.}} \label{fig1}
\end{figure}
\par
Fig. 1 shows the temperature dependence of the pion condensate,
chiral condensate and the trace of Polyakov loop at different
$\mu_I$. In comparison with the zero $\mu_I$ case, the magnitude of
chiral condensate is suppressed by the isospin chemical and it
almost keeps as a constant until the pion condensate evaporates at
high $T_c^{'}$. The dependence of Polyakov loop $\Phi$ on $T$ is
similar to the zero $\mu_I$ case. In the low $\mu_I$ region, the
$I_3$ symmetry restoration phase transition is second order. With
increasing $\mu_I$, the evaporation of the pion condensate gets more
and more abrupt and the $I_3$  symmetry restoration  phase
transition eventually becomes first order. Fig. 1(c) shows that
$\Phi$ is also discontinuous at the first order phase transition
point $T_c^{'}$. This indicates there exists a TCP in the
($T,\mu_I$) phase diagram, which was first proposed in
\cite{Kogut:2002} by using lattice simulation. Fig.1 also shows that
the $I_3$ symmetry restoration phase transition temperature
$T_c^{'}$ is always less then $T_c^{0}$.
\begin{figure}[ht]
\hspace{-.05\textwidth}
\begin{minipage}[t]{.4\textwidth}
\includegraphics*[width=\textwidth]{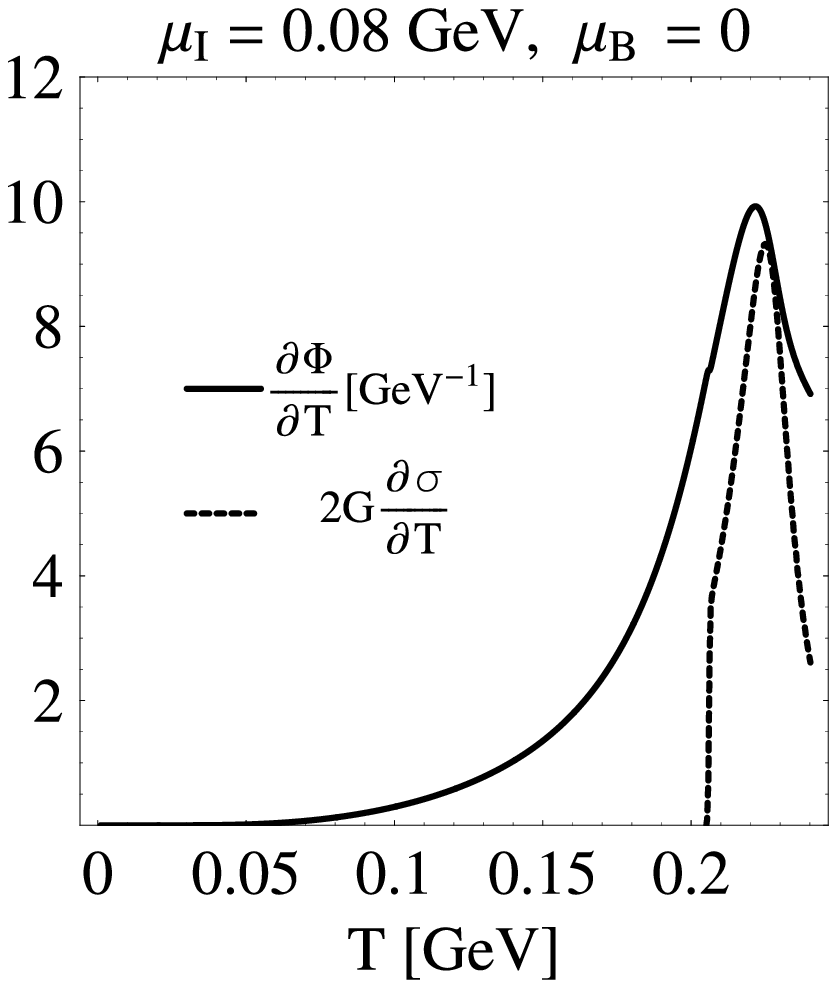}\\
\centerline{(a)}
\end{minipage}
\hspace{.05\textwidth}
\begin{minipage}[t]{.4\textwidth}
\hspace{-.05\textwidth} \scalebox{.85}{
\includegraphics*[width=1.17\textwidth]{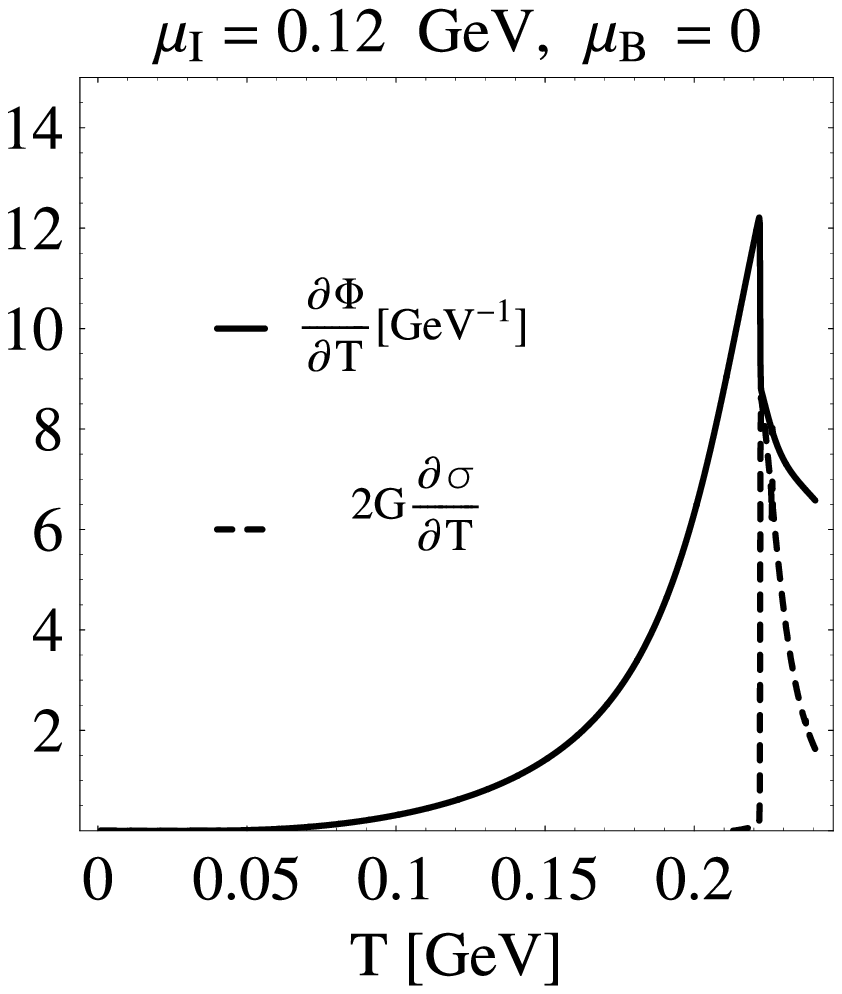}}\\
\centerline{(b)}
\end{minipage}
\centering
\hspace{-.05\textwidth}
\begin{minipage}[t]{.4\textwidth}
\hspace{-.10\textwidth}\scalebox{.85}{
\includegraphics*[width=1.17\textwidth]{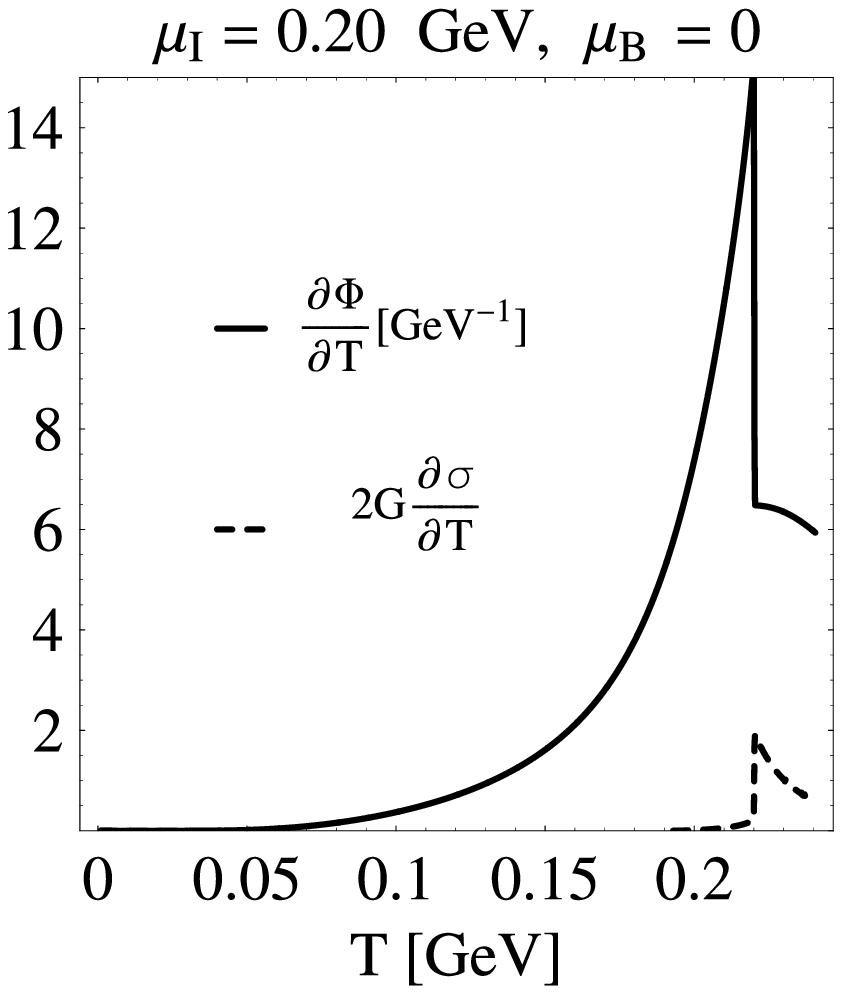}}\\
\centerline{(c)}
\end{minipage}
\parbox{15cm}{
\caption{Plots of $\partial({2G\sigma})/\partial T$ and
$\partial\Phi/
\partial T$ for different $\mu_I$ with $\lambda=0$.
} \label{fig2}}
\end{figure}
\par
Although chiral condensate is suppressed by the pion condensate, it
still keeps considerable value in the low $\mu_I$ region with
$\mu_I>\mu_c$. Fig. 2 shows that the crossover phase transitions for
the chiral condensate $\sigma$ and for the Polyakov loop $\Phi$ also
perfectly coincide at $T_c$ with $T_c^{'}<T_c<T_c^{0}$. Therefore,
if we define the peak of the Polyakov loop
${\text{susceptibility}}^{1}$\footnotetext[1]{In this paper the
Polyakov loop susceptibility is defined as the partial differential
of $\Phi$ with respect to $T$ or $\mu_I$. The definitions of the
susceptibilities for chiral condensate and pion condensate are
alike. } as the deconfinement phase transition point, the pion
superfluidity phase is always confined at low $\mu_I$ region. For
high $\mu_I$, chiral condensate is greatly reduced and it has no
meaning to compare the chiral susceptibility with the Polyakov loop
susceptibility. Numerical results indicate that the peak of Polyakov
loop susceptibility excellently  coincides with the pion condensate
vanishing point for high enough $\mu_I$, while for the not too large
$\mu_I$ the difference of these two points is within a few
$\mathrm{MeV}$. This points will be illustrated more clearly in the
next section by considering a small $I_3$ symmetry breaking term.
Therefore, we get conclusion that the pion superfluidity phase is
always confined.
\par
\begin{figure}[ht]
\hspace{-.05\textwidth}
\begin{minipage}[t]{.4\textwidth}
\includegraphics*[width=\textwidth]{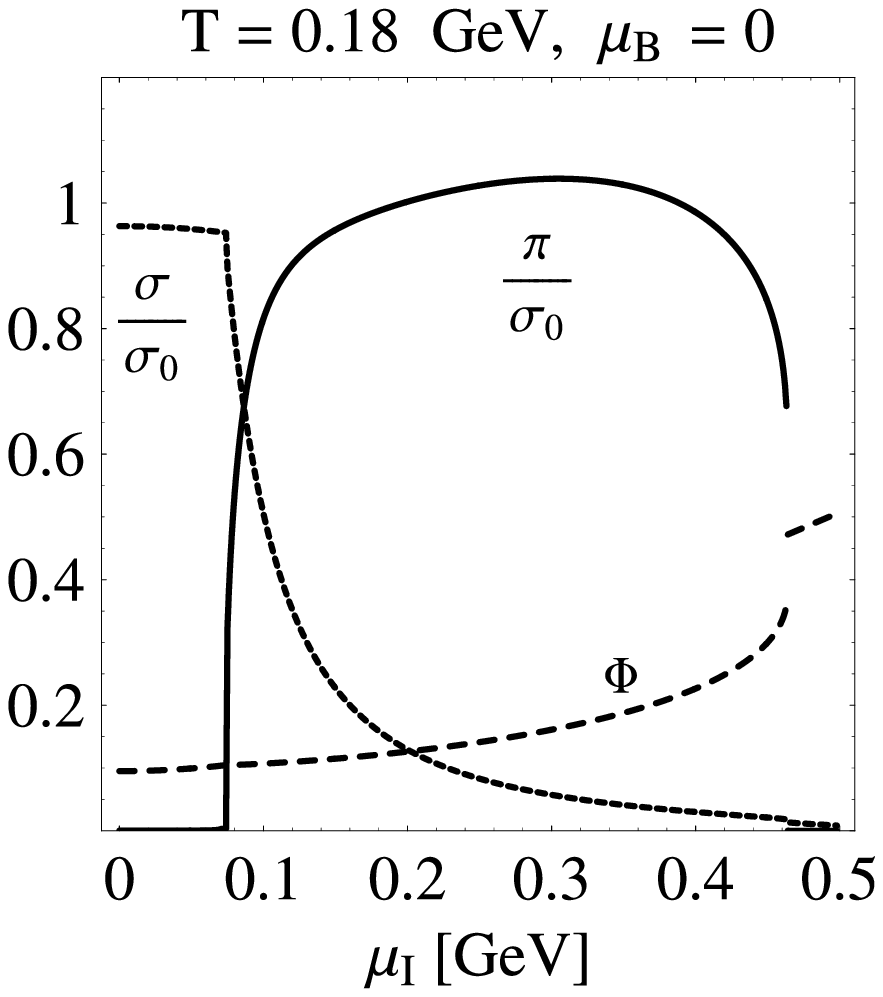}\\
\centerline{(a)}
\end{minipage}
\hspace{.05\textwidth}
\begin{minipage}[t]{.4\textwidth}
\hspace{-.05\textwidth} \scalebox{.86}
{\includegraphics*[width=1.17\textwidth]{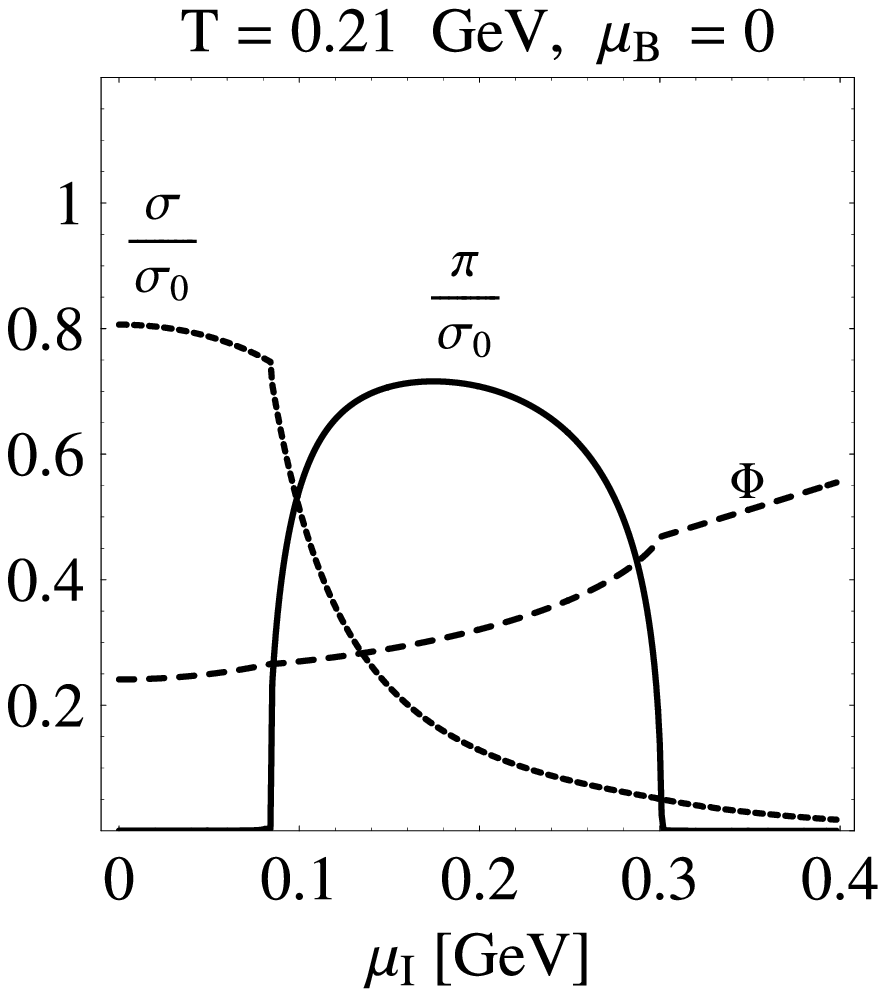}}\\
\centerline{(b)}
\end{minipage}
\centering
\hspace{-.05\textwidth}
\begin{minipage}[t]{.4\textwidth}
\hspace{-.10\textwidth} \scalebox{.90}{
\includegraphics*[width=1.17\textwidth]{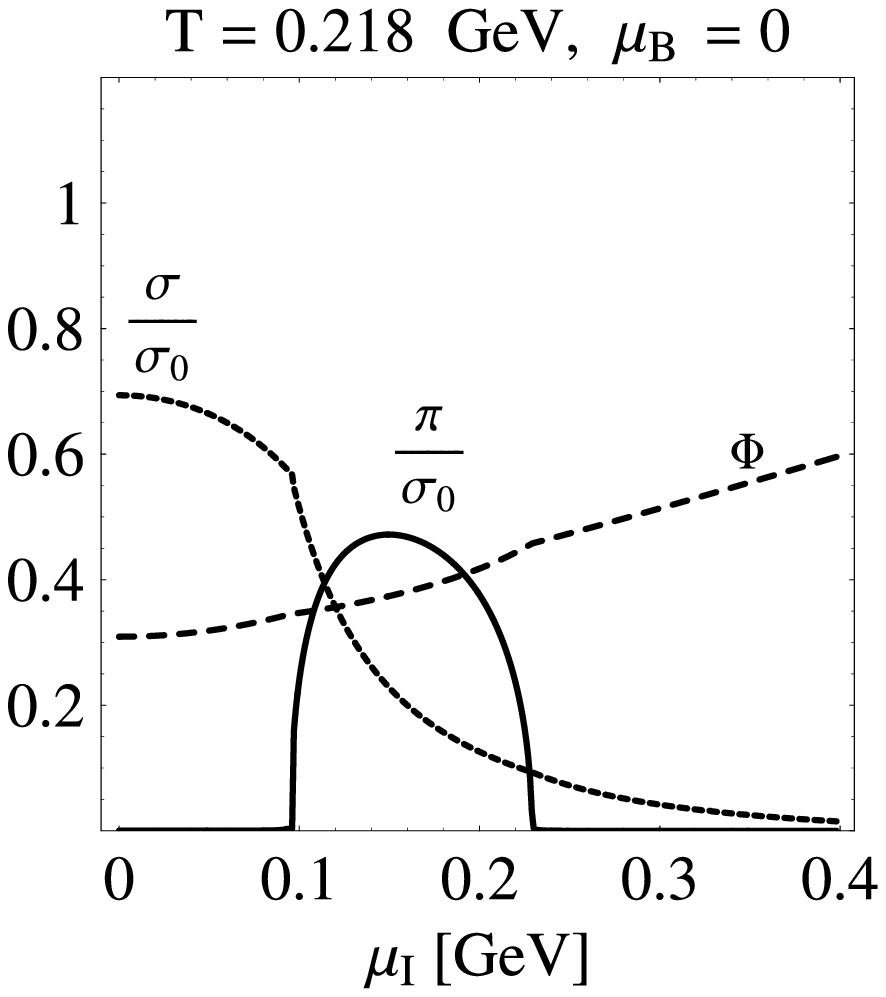}}\\
\centerline{(c)}
\end{minipage}
\parbox{15cm}
{\caption{Scaled pion condensate, chiral condensate and Polyakov
loop $\Phi$ as functions of isospin chemical potential at zero
baryon chemical potential for different $T$ with $\lambda=0$.}
 \label{fig3}}
\end{figure}
The isospin chemical dependence of $\sigma$, $\pi$ and $\Phi$ are
plotted in Fig. 3.  For a fixed $T$,  the Polyakov loop $\Phi$ is
always a monotonic increasing function of $\mu_I$ and the scaled
chiral condensate is always a  monotonic decreasing function of
$\mu_I$. These points also qualitatively coincide with the lattice
results \cite{Kogut:2002,Kogut:2004}. For $T=180\mathrm{MeV}$, pion
superfluidity phase occurs at $\mu_I=75\mathrm{MeV}$ which is very
close to $\mu_c$ and the phase transition is  second order. The
scaled pion condensate first increases with the increasing $\mu_I$
and then decreases with the increasing $\mu_I$ until it abruptly
vanishes at $\mu_I=456\mathrm{MeV}$. With increasing $T$, the first
order phase transition for the pion condensate vanishing becomes
more and more soft and it evolves as a second order phase transition
when $T>210\mathrm{MeV}$. Fig. 3 also shows that the two phase
transition points for $I_3$ symmetry breaking and restoration get
more and more close to each other with the increasing $T$ and
eventually coincide at $\sim$($223\mathrm{MeV},135\mathrm{MeV}$).
Numerical results also suggest that the peak of
$\partial{\Phi}/\partial{\mu_I}$ is located at the vicinity of the
pion condensate vanishing point.
\par
The reason that the pion condensate first increases and then
decreases with the increasing $\mu_I$ at fixed $T$ is that it
eventually becomes energetically favourable to produce fermionic
excitations which then competes with the production of fermions in
bound states with anti-fermions, that is pions. It is reasonable
that to produce quark excitations gets more and more advantaged in
contrast to the production of quarks in pions with anti-quarks for
high $T$, as suggested by Fig.3. Since this phenomena is also
observed at $T\sim223\mathrm{MeV}$ where the non vanishing pion
condensate(with very small value) only appears in the vicinity of
$\mu_I\sim135\mathrm{MeV}$, we believe that the above result is not
an effect due to the ultraviolet cutoff.
\par
\begin{figure}[ht]
\hspace{-.05\textwidth}
\begin{minipage}[t]{.45\textwidth}
\includegraphics*[width=\textwidth]{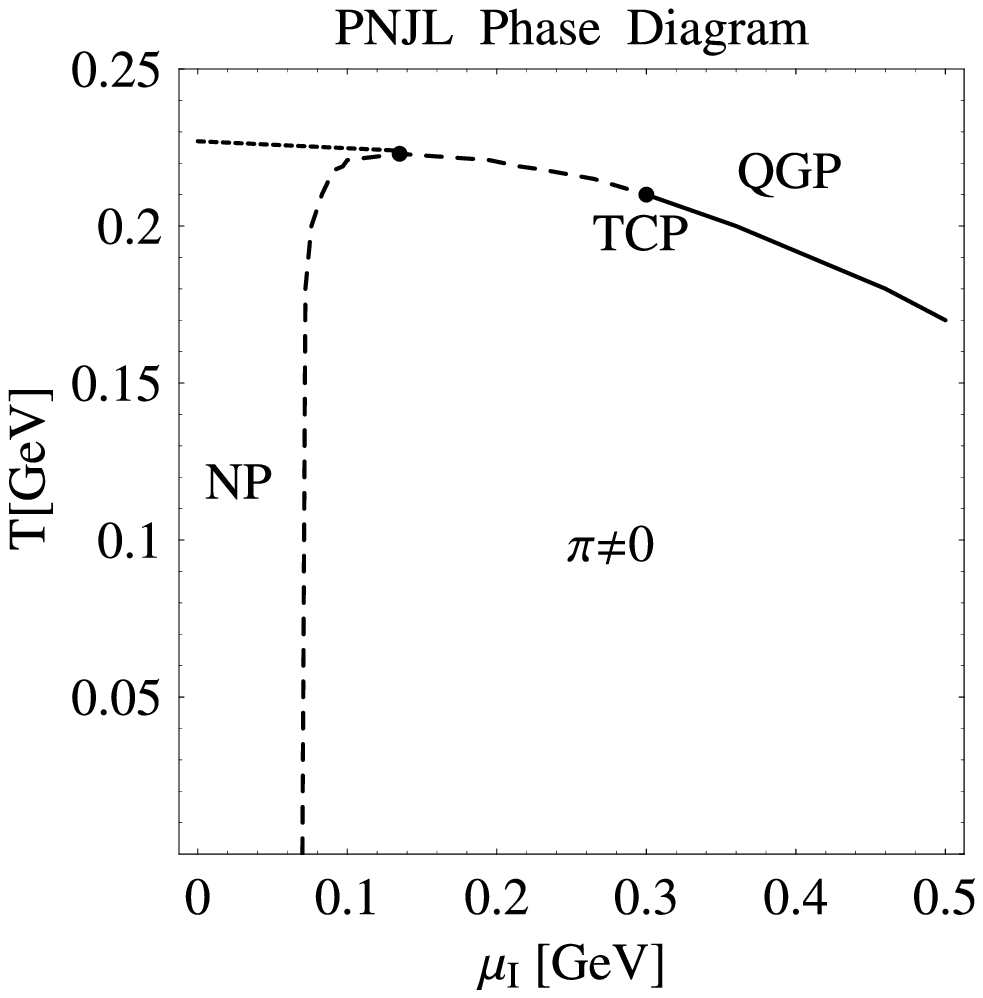}\\
\centerline{(a)}
\end{minipage}
\hspace{-.05\textwidth}
%
\begin{minipage}[t]{.45\textwidth}
\hspace{-.15\textwidth} \scalebox{.85}{
\includegraphics*[width=1.17\textwidth]{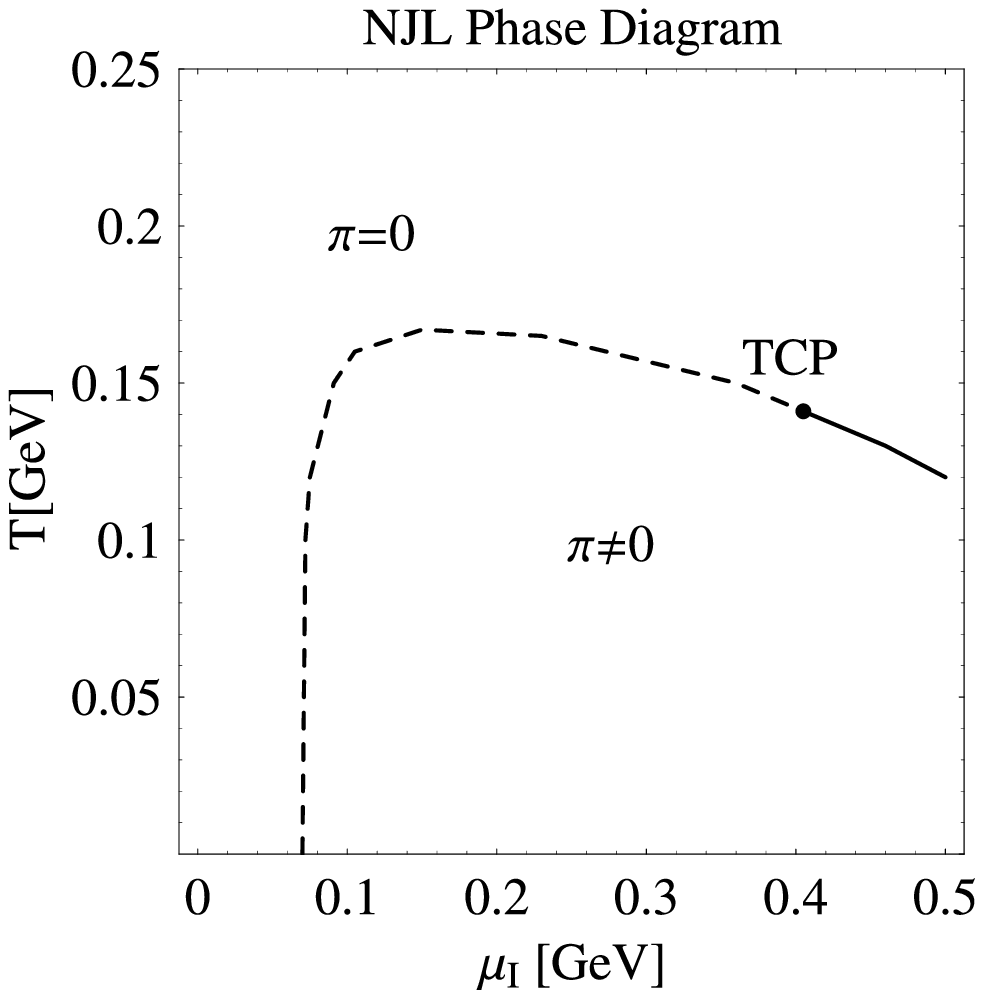}}\\
\centerline{(b)}
\end{minipage}
\parbox{15cm}{
\caption{The ($T,\mu_I$) phase diagrams of two flavor PNJL (Left)
and standard NJL (Right). NP stands for normal hadronic phase and
QGP stands for quark-gluon plasma phase. The solid line (dashed
line) stands for the first order ( second order ) pion superfluidity
phase transition. The dotted line in the left panel indicates the
crossover for deconfinement phase transition.}
 \label{fig4}}
\end{figure}
The ($T,\mu_I$) phase diagram of two flavor PNJL model is shown in
Fig. 4(a), in comparison with the standard NJL phase diagram
described  by Fig. 4(b). We only plots the phase diagrams in the
range $\mu_I<500\mathrm{MeV}$ since the investigation for more high
isospin chemical potential is beyond the capability of the NJL type
models. The PNJL phase diagram is qualitatively consistent with the
lattice results given in \cite{Kogut:2004}. Fig. 4(a) shows that the
pion condensate always vanishes  above $T_c^{0}$. The TCP is located
at $(210\mathrm{MeV},300\mathrm{MeV})$, which means that the phase
transition for $I_3$ symmetry restoration in  the region
$\mu_I>300\mathrm{MeV}$ is always first order.  In the rough, for
$\mu_I>135\mathrm{MeV}$, the chiral condensate is greatly suppressed
and the melting of pion condensate is simultaneously accompanied by
the crossover phase transition for deconfinement; while for
$\mu_c<\mu_I<135\mathrm{MeV}$, the chiral condensate still keeps a
considerable value when the pion condensate vanishes and the
crossover phase transitions for chiral symmetry restoration and
deconfinement are still coincide. Note that numerical study suggests
that the standard NJL model also sustain the existence of the TCP in
the ($T,\mu_I$) phase diagram. The TCP obtained in standard NJL is
at the vicinity of $(140\mathrm{MeV},400\mathrm{MeV})$. However, the
standard NJL has no capability to give the information concerning
the deconfinement phase transition.
\par
\begin{figure}[ht]
\hspace{-.05\textwidth}
\begin{minipage}[t]{.4\textwidth}
\includegraphics*[width=\textwidth]{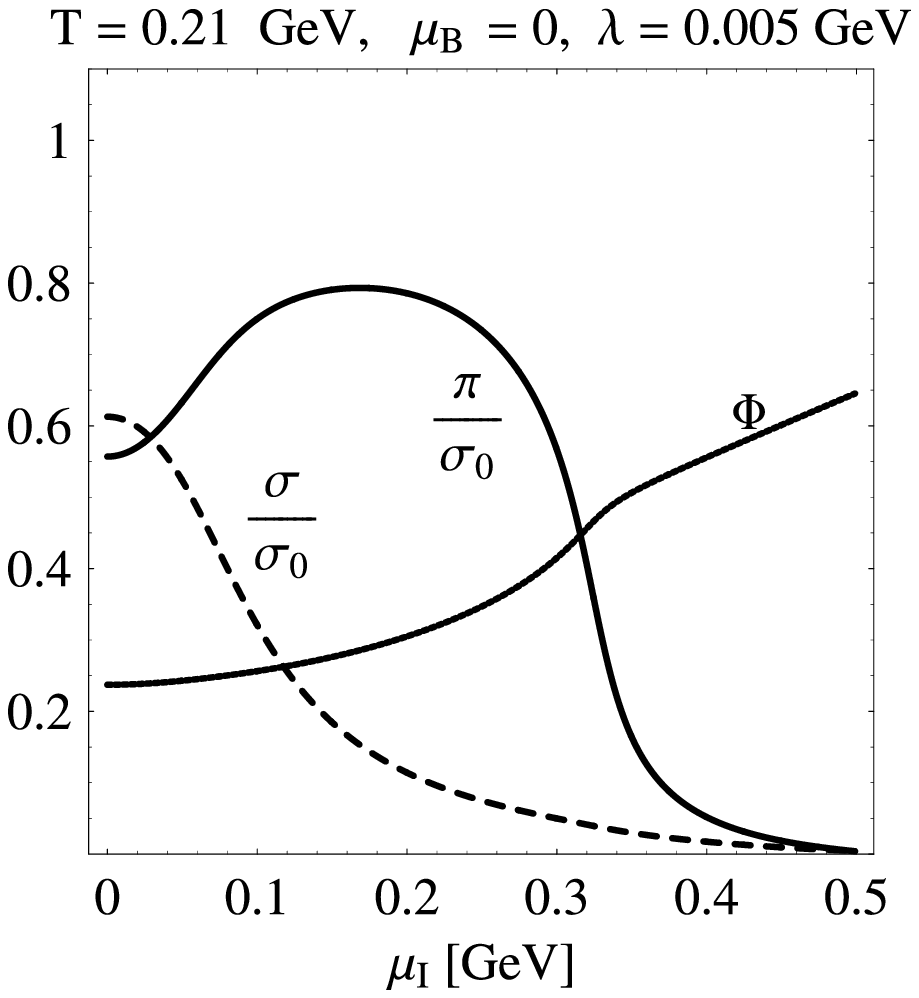}\\
\centerline{(a)}
\end{minipage}
\hspace{.05\textwidth}
\begin{minipage}[t]{.4\textwidth}
\hspace{-.05\textwidth} \scalebox{.87}
{\includegraphics*[width=1.17\textwidth]{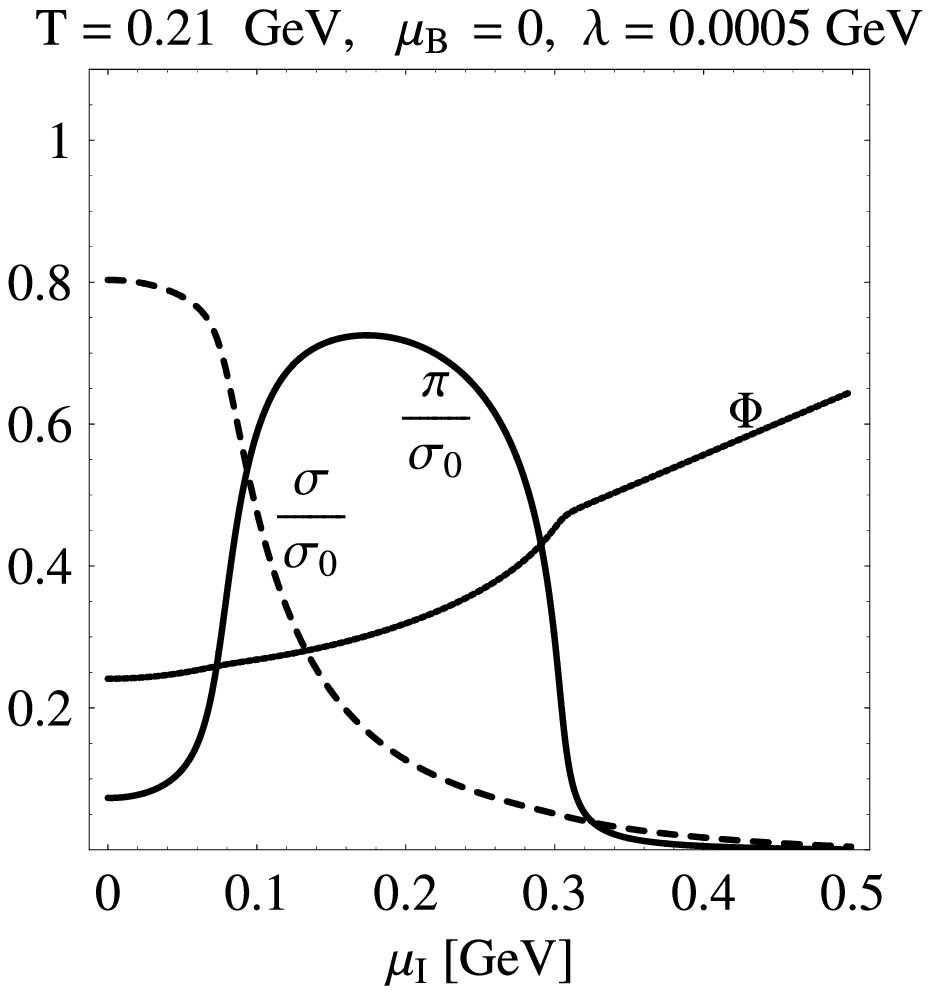}}\\
\centerline{(b)}
\end{minipage}
\centering \hspace{-.05\textwidth}
\begin{minipage}[t]{.4\textwidth}
\hspace{-.10\textwidth} \scalebox{.90}{
\includegraphics*[width=1.17\textwidth]{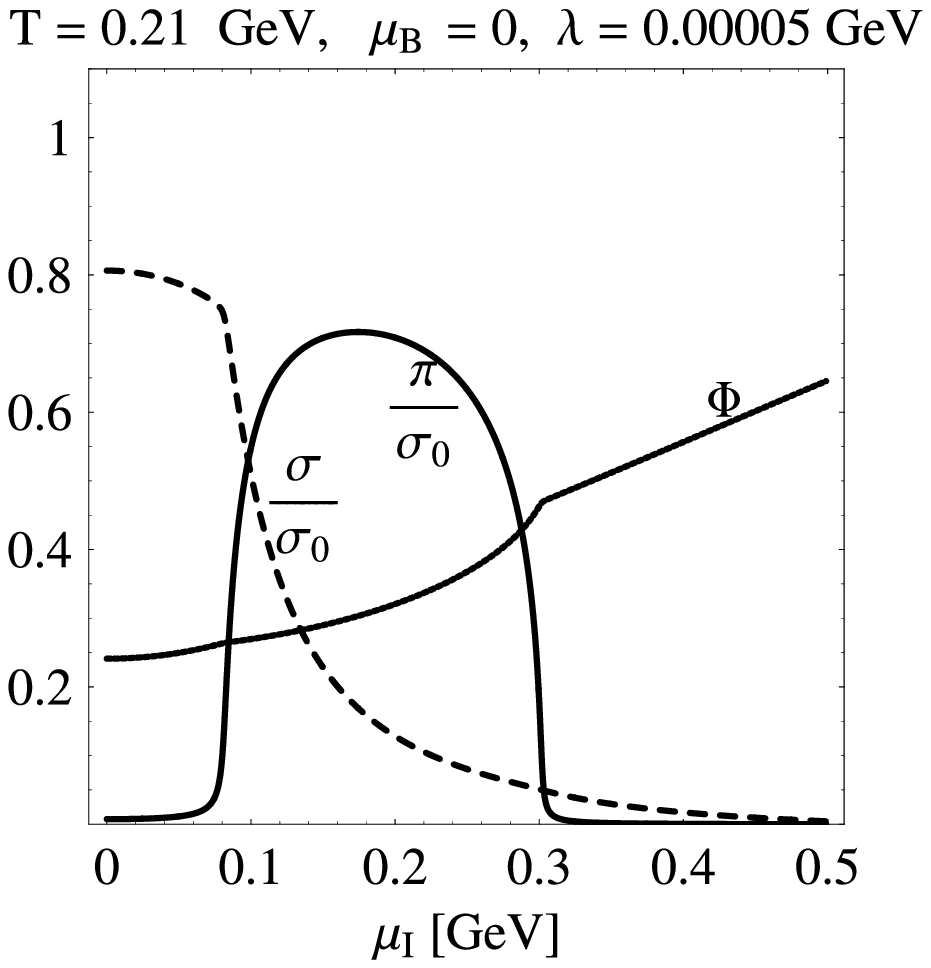}}\\
\centerline{(c)}
\end{minipage}
\parbox{15cm}
{\caption{Scaled pion condensate, chiral condensate and Polyakov
loop $\Phi$ as functions of isospin chemical potential at zero
baryon chemical potential for $T=210\mathrm{MeV}$ with different
$\lambda\neq0$.}
 \label{fig5}}
\end{figure}

\subsection{The case for $\lambda\neq{0}$ and $\mu_B=0$}
The previous lattice simulations for pion superfluidity phase
transition are performed by introducing an explicit isospin symmetry
term which is proportional to a small parameter $\lambda$ in the
lagrangian.  The phase diagram is then obtained by extrapolating the
results from $\lambda\neq{0}$ to $\lambda=0$. To compare with the
lattice data, we also investigate the $\lambda\neq{0}$ within PNJL
formalism. In addition, studying $\lambda\neq{0}$ case can further
deepen our understanding the role of the Polyakov loop dynamics .
\par
\begin{figure}[ht]
\hspace{-.05\textwidth}
\begin{minipage}[t]{.4\textwidth}
\includegraphics*[width=\textwidth]{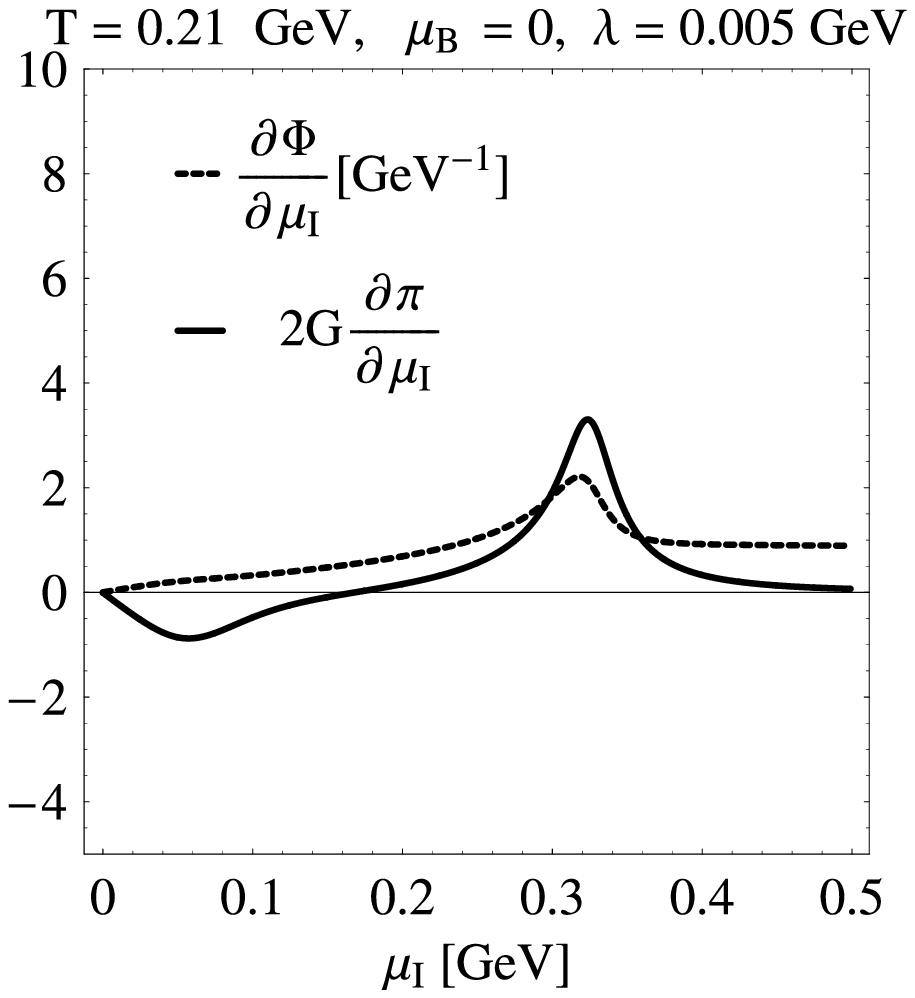}\\
\centerline{(a)}
\end{minipage}
\hspace{0.05\textwidth}
\begin{minipage}[t]{.4\textwidth}
\hspace{-.05\textwidth} \scalebox{.86}
{\includegraphics*[width=1.17\textwidth]{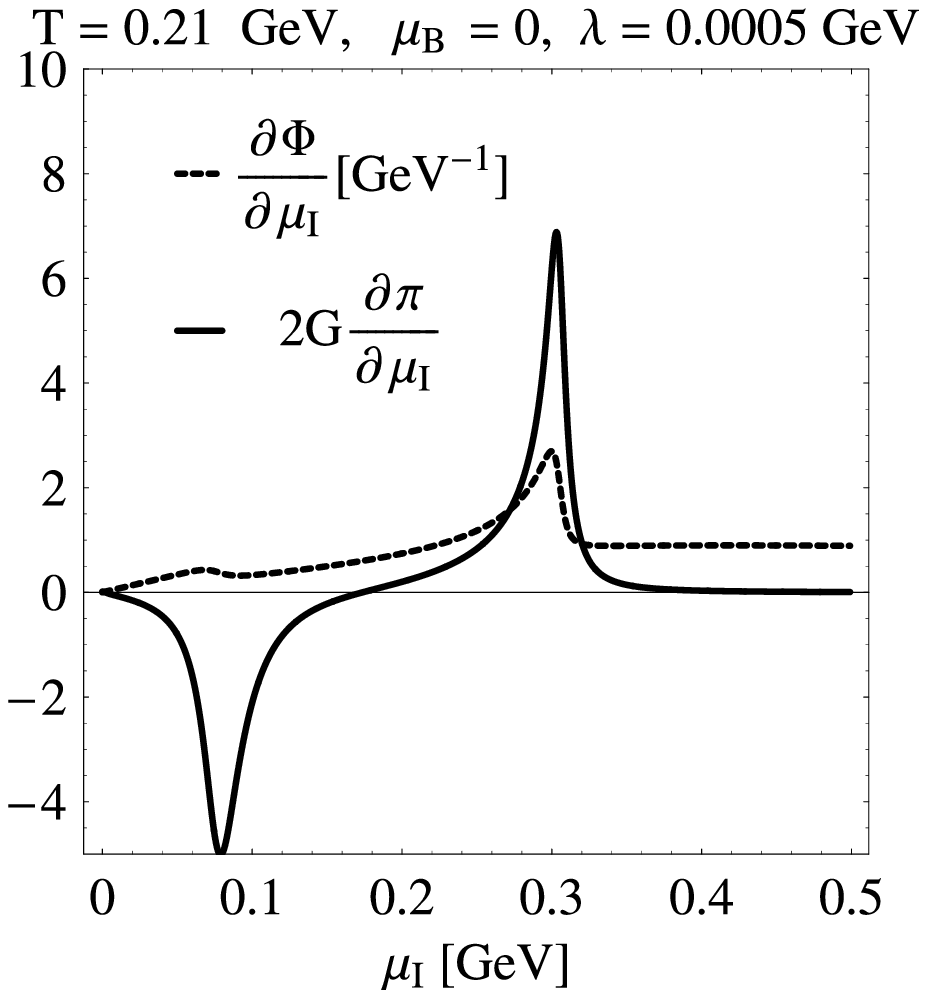}}\\
\centerline{(b)}
\end{minipage}
\centering \hspace{-.05\textwidth}
\begin{minipage}[t]{.4\textwidth}
\hspace{-.10\textwidth} \scalebox{.93}{
\includegraphics*[width=1.17\textwidth]{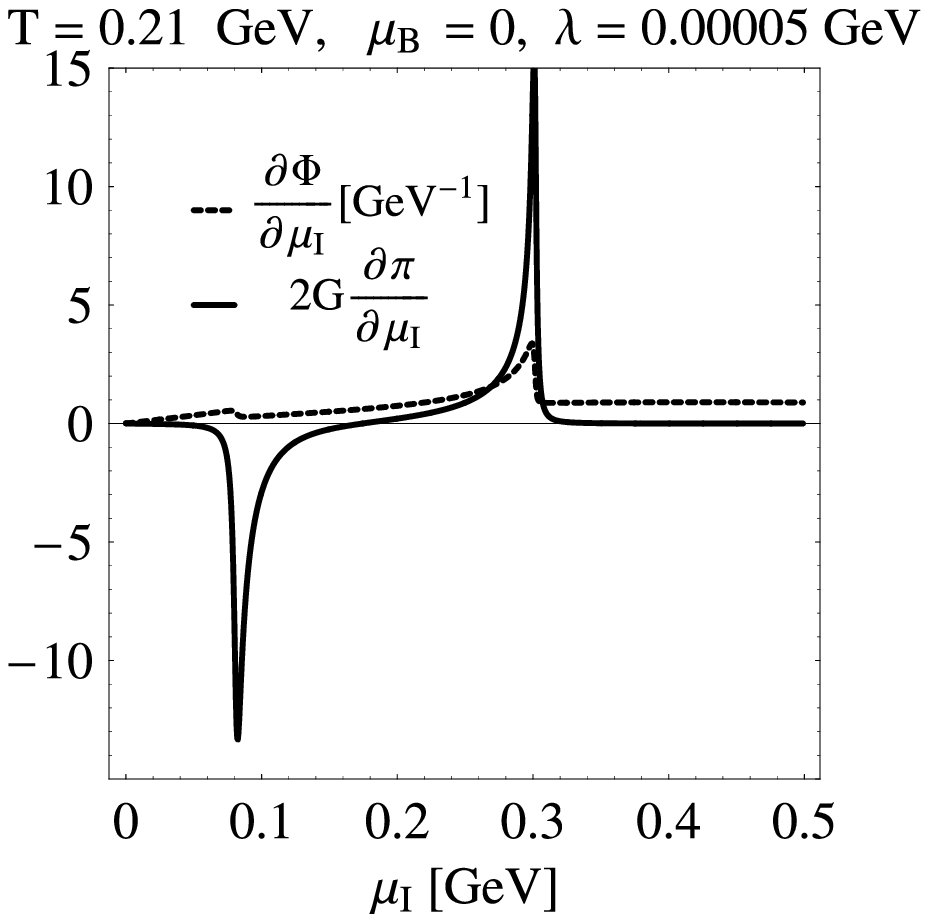}}\\
\centerline{(c)}
\end{minipage}
\parbox{15cm}
{\caption{Plots of $\partial({2G\pi})/\partial\mu_I$ and
$\partial\Phi/
\partial\mu_I$  as functions of $\mu_I$ for $T=210\mathrm{MeV}$ and $\mu_B=0$ with different
$\lambda$.}
 \label{fig6}}
\end{figure}
with $\lambda\neq{0}$, the pion condensate is no longer the true
order parameter and the phase transition for pion superfluidity
becomes a crossover. We perform the calculations with
$\lambda=5\mathrm{MeV}$(with the same order as $m_0$),
$0.5\mathrm{MeV}$ and $0.05\mathrm{MeV}$, respectively. Fig. 5 gives
the scaled pion condensate, chiral condensate and Polyakov loop as
functions of isospin chemical potential at $\mu_B=0$ and
$T=210\mathrm{MeV}$ for different $\lambda$. Fig. 5(a) shows that
when $\lambda=m_0$ the chiral condensate and pion condensate will
have the same magnitude in vacuum, which indicates that a small
explicit symmetry term has significant influence on the vacuum
structure. With $\lambda$ decreasing, the magnitude of $\pi$ in
$\mu_I<\mu_c$ is greatly suppressed while its influence on the large
$\mu_I$ region is not so significantly. We can see that Fig. 5(c)
which is obtained with $\lambda=0.05\mathrm{MeV}$ is almost
identical with Fig. 3(b) which is  obtained with $\lambda=0$. Fig. 5
also shows varying of small $\lambda$ has little impact on the shape
of Polyakov loop $\Phi$. Here we can get conclusion that the method
taken in lattice simulation by extrapolating the results from
$\lambda\neq{0}$ to $\lambda=0$ is reliable.
\par
\begin{figure}[ht]
\hspace{-.05\textwidth}
\begin{minipage}[t]{.4\textwidth}
\includegraphics*[width=\textwidth]{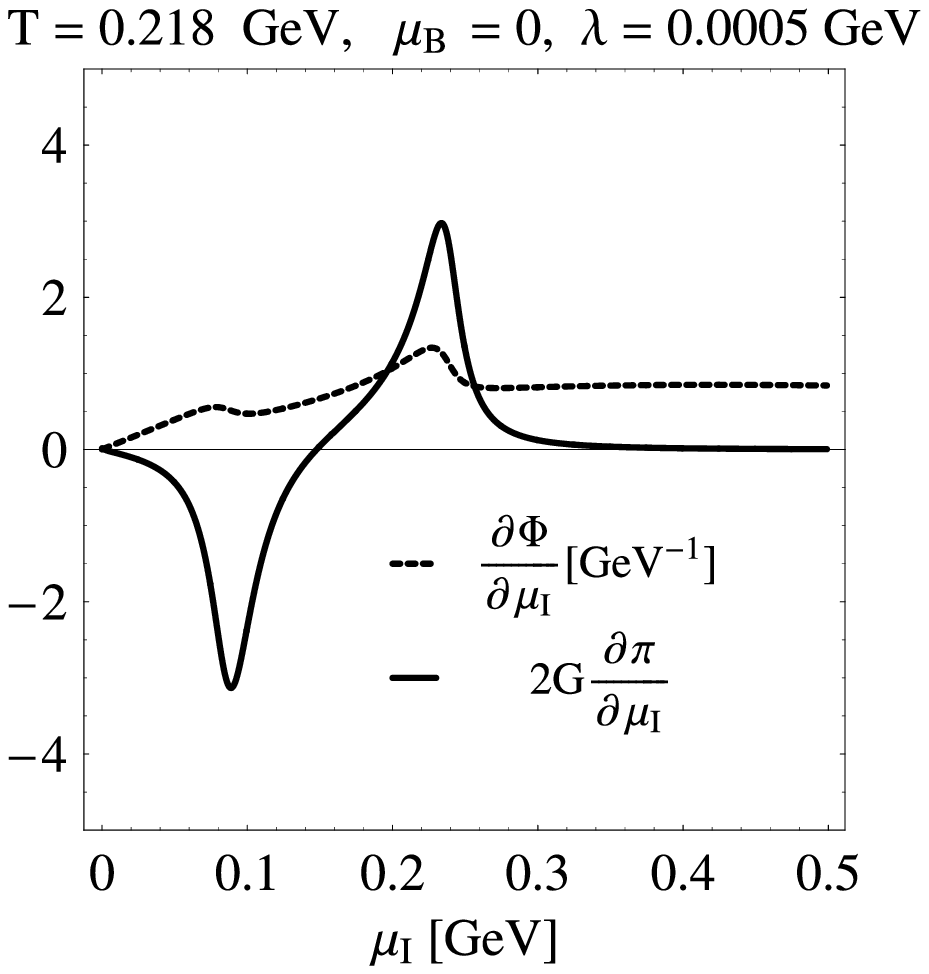}\\
\centerline{(a)}
\end{minipage}
\hspace{.05\textwidth}
\begin{minipage}[t]{.4\textwidth}
\hspace{-.05\textwidth} \scalebox{.84}
{\includegraphics*[width=1.17\textwidth]{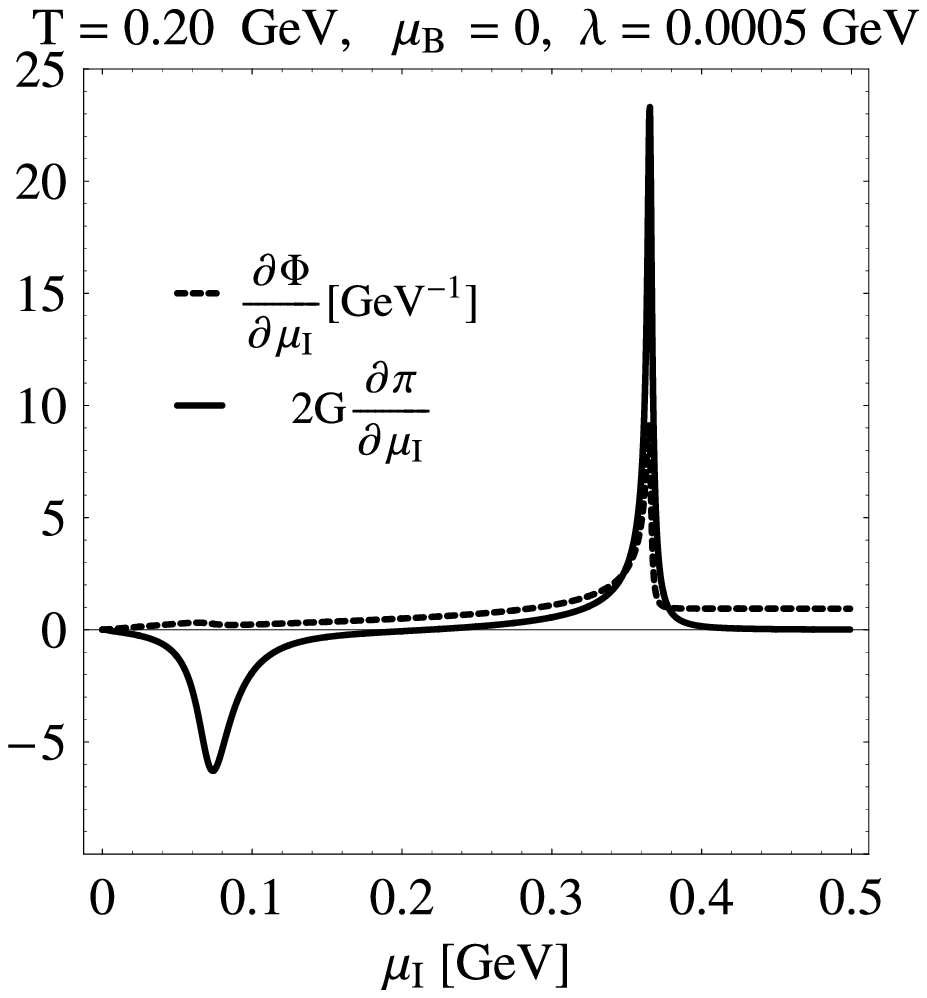}}\\
\centerline{(b)}
\end{minipage}
\centering \hspace{-.05\textwidth}
\begin{minipage}[t]{.4\textwidth}
\hspace{-.10\textwidth} \scalebox{.88}{
\includegraphics*[width=1.17\textwidth]{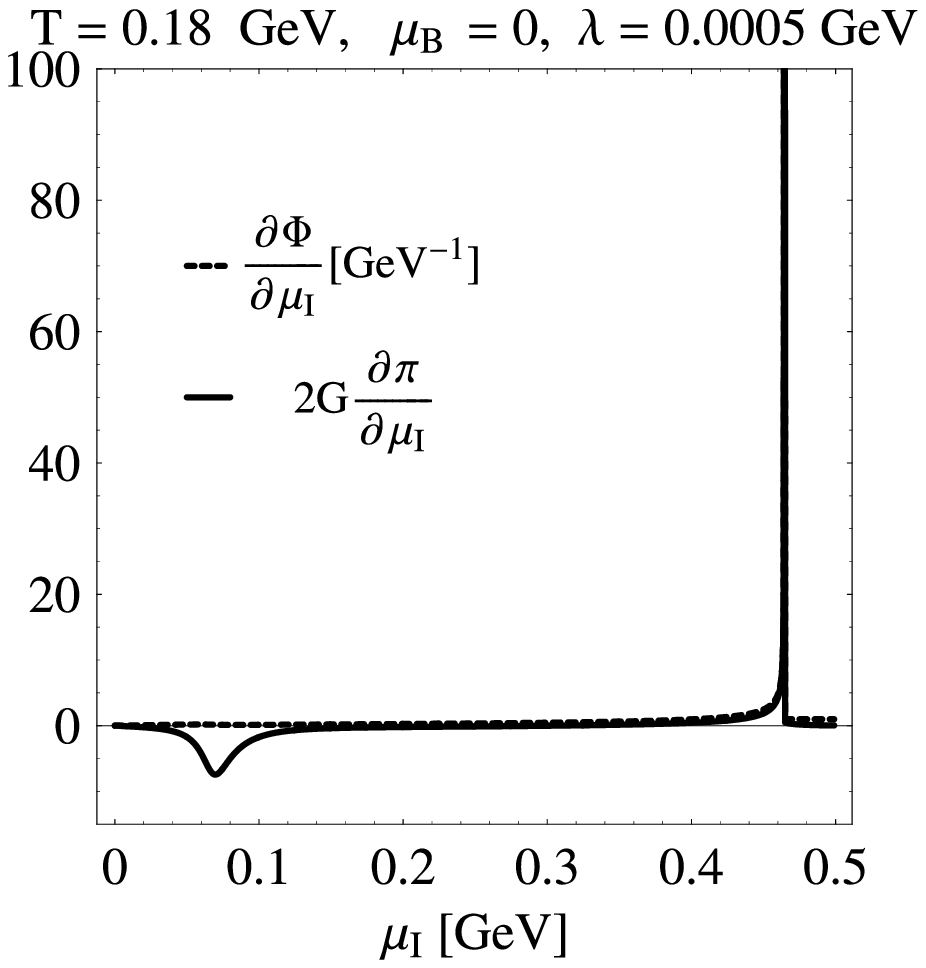}}\\
\centerline{(c)}
\end{minipage}
\parbox{15cm}
{\caption{Plots of $\partial({2G\pi})/\partial\mu_I$ and
$\partial\Phi/
\partial\mu_I$ as functions of $\mu_I$ at $\lambda=0.5\mathrm{MeV}$
for different temperature.}
 \label{fig7}}
\end{figure}
For small $\lambda$, we can use the peak of the pion condensate
susceptibility to indicate the crossover phase transition for $I_3$
symmetry breaking or restoration. In Fig. 6, we present
$\partial{\pi}/\partial\mu_I$ and $\partial\Phi/
\partial\mu_I$  obtained  with the same parameters as Fig. 5.
The curve of $\partial{\pi}/\partial\mu_I$ has two peaks, which
 correspond to  the crossover phase transitions for $I_3$ symmetry breaking
 (the left peak) and $I_3$ symmetry restoration (the right peak), respectively.
  It is shown that the peak of $\partial\Phi/\partial\mu_I$ perfectly coincides with the right
peak of $\partial{\pi}/\partial\mu_I$. Fig. 6 also shows that the
peak position of $\partial\Phi/\partial\mu_I$ is insensitive to the
value of $\lambda$ when $\lambda<<m_0$ while the extent of
superposition between the peaks of $\partial\Phi/\partial{\mu_I}$
and $\partial{\pi}/\partial{\mu_I}$ is closely related to the
magnitude of the explicit symmetry breaking term. In comparison with
the Fig. 3(b), we find that the peak position of crossover phase
transition for $I_3$ symmetry breaking  or restoration is
considerably consistent with the corresponding true phase transition
point for $I_3$ symmetry breaking or restoration obtained with
$\lambda=0$. This results confirm that the melting of pion
condensate for $\mu_I>\mu_c^{'}\approx135\mathrm{MeV}$ is
simultaneously accompanied by the crossover deconfinement phase
transition, just as shown in Fig. 4(a).
\par
\begin{figure}[ht]
\hspace{-.05\textwidth}
\begin{minipage}[t]{.4\textwidth}
\includegraphics*[width=\textwidth]{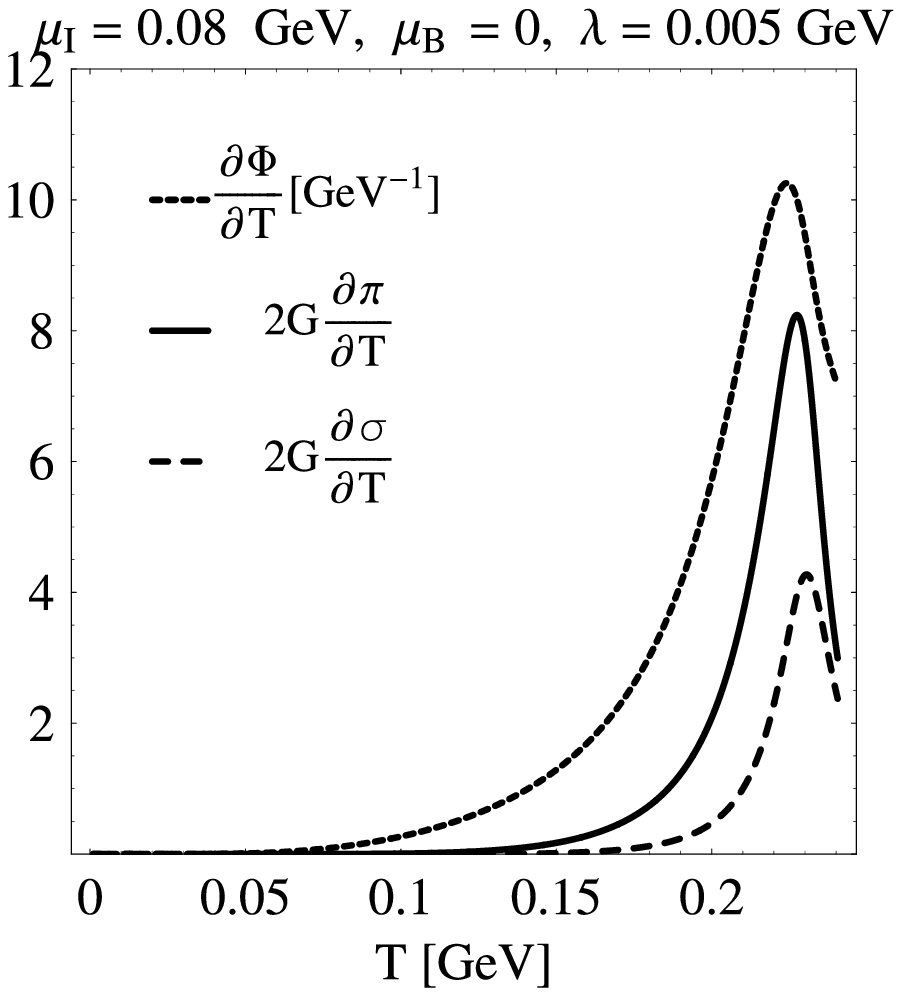}\\
\centerline{(a)}
\end{minipage}
\hspace{.05\textwidth}
\begin{minipage}[t]{.4\textwidth}
\hspace{-.05\textwidth} \scalebox{.85}
{\includegraphics*[width=1.17\textwidth]{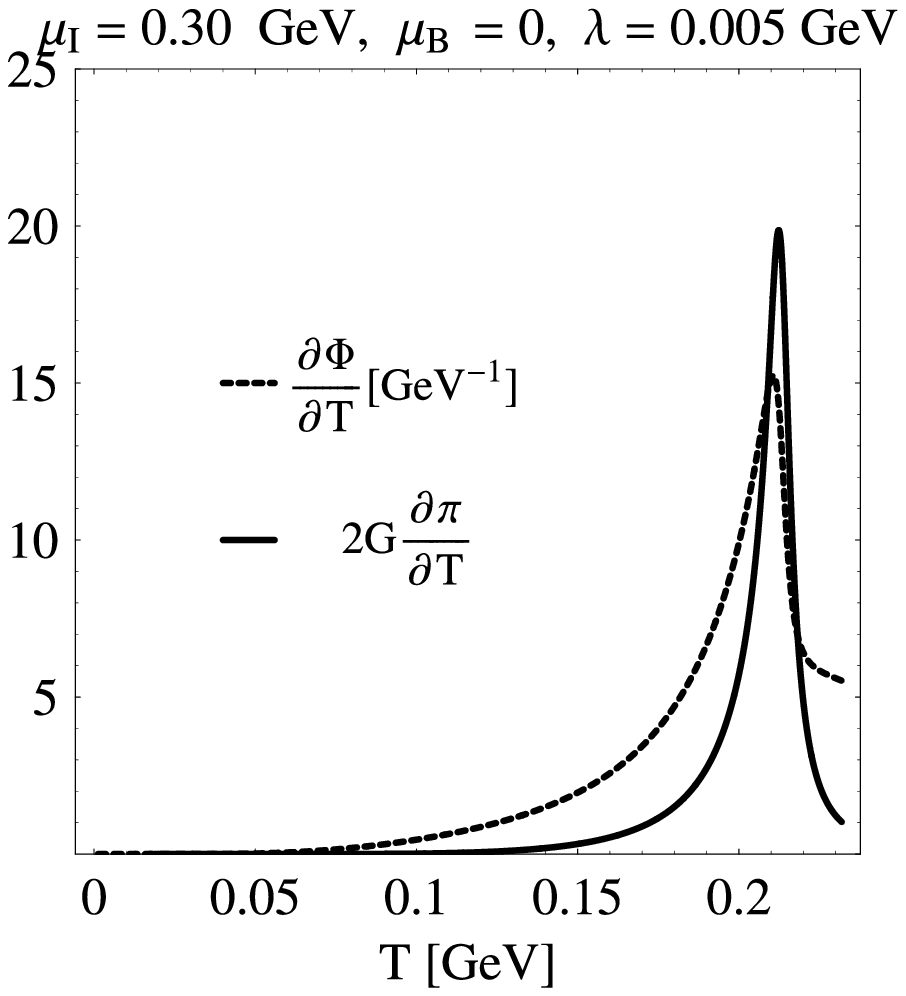}}\\
\centerline{(b)}
\end{minipage}
\centering \hspace{-.05\textwidth}
\begin{minipage}[t]{.4\textwidth}
\hspace{-.10\textwidth} \scalebox{.90}{
\includegraphics*[width=1.17\textwidth]{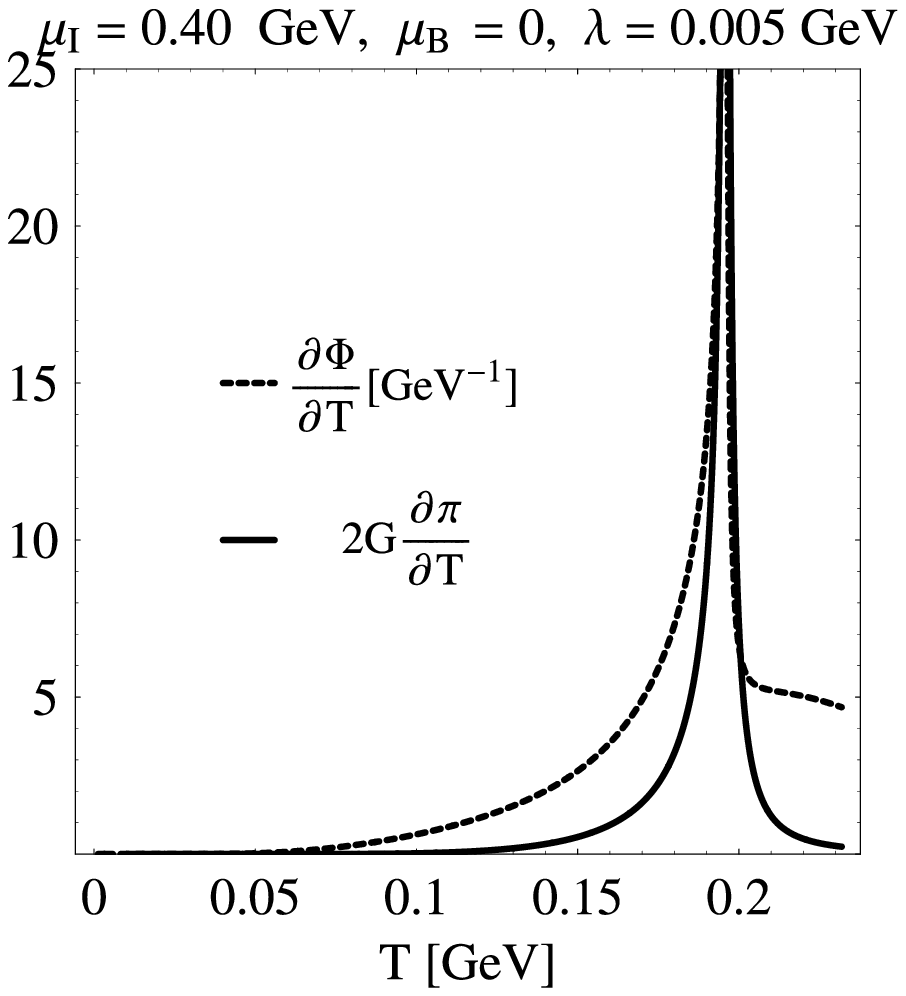}}\\
\centerline{(c)}
\end{minipage}
\parbox{15cm}
{\caption{Plots of $\partial({2G\sigma})/\partial{T}$,
$\partial({2G\pi})/\partial{T}$ and $\partial\Phi/
\partial{T}$  as functions of T  for different $\mu_I$  with the
same $\lambda$.}
 \label{fig8}}
\end{figure}
The evidence for the existence of TCP is illustrated in Fig. 7.
Fixing $\lambda=0.5\mathrm{MeV}<<m_0$, for high enough
temperature($T>210\mathrm{MeV}$), the crossover phase transition for
$I_3$ symmetry breaking is similar to the crossover phase transition
for $I_3$ symmetry restoration, which indicates that both phase
transitions are second orders for $\lambda=0$; While for the low
enough temperature($T<210\mathrm{MeV}$), the right peak becomes
significantly steeper than the left one, which indicates that it
corresponds to a first order phase transition for $\lambda=0$. Fig.
7 also shows that  the two peaks get more and more close to each
other with increasing $T$ and eventually coincide at a point
$(T^{'},\mu^{'})$. For  $\lambda<<m_0$, we find that the point
$(T^{'},\mu^{'})$ is almost identical with the point
($223\mathrm{MeV},135\mathrm{MeV}$).
\par
Above results can also be confirmed by exploring the peak positions
of $\partial{\pi}/\partial{T}$ and $\partial\Phi/
\partial{T}$, which is shown in Fig. 8. For low
$\mu_I=80\mathrm{MeV}$, we can see that differences among three
crossover phase transitions for chiral condensate, pion condensate
and Polyakov loop are still within a few $\mathrm{MeV}$ with
$\lambda\approx{m_0}$. For the same $\lambda$,  the peaks related to
the pion condensate and Polyakov loop both get more and more steep
with increasing $\mu_I$, which implies that second order phase
transition for $I_3$ symmetry restoration eventually involves into
the first order phase transition.
\begin{figure}[ht]
\hspace{-.05\textwidth}
\begin{minipage}[t]{.45\textwidth}
\includegraphics*[width=\textwidth]{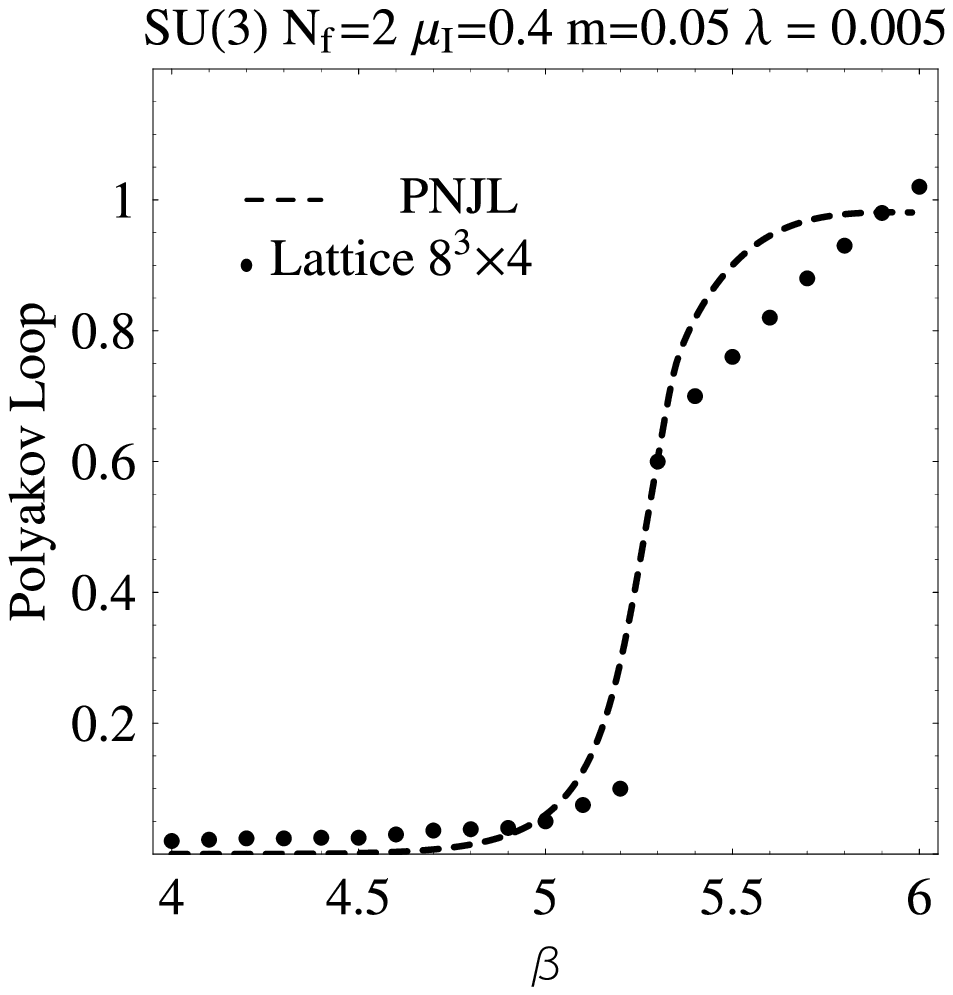}\\
\centerline{(a)}
\end{minipage}
\hspace{-.08\textwidth}
\begin{minipage}[t]{.45\textwidth}
\hspace{-.05\textwidth} \scalebox{0.8}
{\includegraphics*[width=1.17\textwidth]{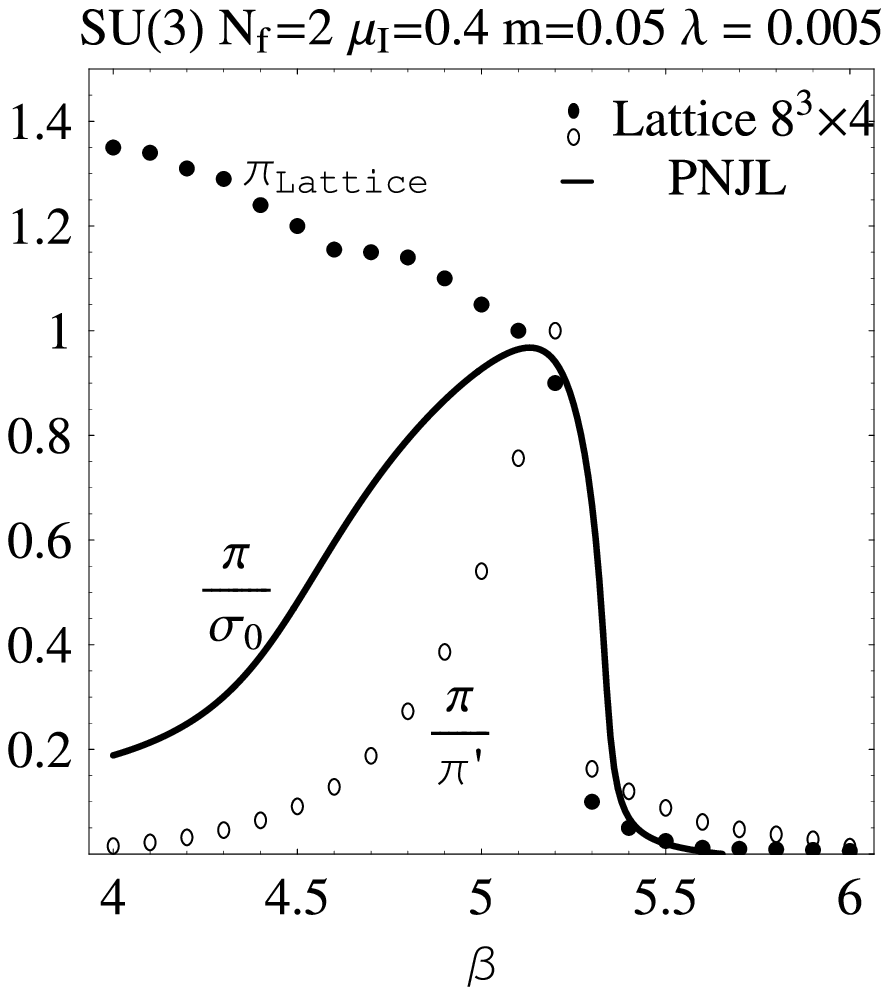}}\\
\centerline{(b)}
\end{minipage}
\parbox{15cm}
{\caption { Polyakov Loop and pion condensate as a function of
$\beta$. The lattice data (the black circles) are taken from
\cite{Kogut:2002} for $m=0.05, \lambda=0.005$ on an $8^{3}\times4$
lattice. In the right panel,  the white circles correspond to the
original lattice data which  were first represented in the same
physical units and then rescaled by the obtained pion condensate at
$\beta=5.2$($\pi'$).  }} \label{fig9}
\end{figure}
\subsection{Comparison with Lattice results}
For quantitative comparison with the corresponding lattice data,
following \cite{Ratti1}, we also reduce the $T_c^0$ by rescaling the
parameter $T_0$ from $270\text{MeV}$ to $190\text{MeV}$. In this
case, though the perfect coincidence of the crossovers for the
deconfinement and chiral phase transitions is lost, the difference
between the two transition  is within $20\text{MeV}$\cite{Ratti1}.
The average of these two temperatures can be defined as $T_c^0$ with
the value $180\text{MeV}$, which is in the range of recent lattice
results\cite{rbcBI2+1,fodorTc}. Note that in the following
calculations, we keep the liquid quark mass varying with the change
of $\beta$ and other parameters of the PNJL unchanged.
\par
The PNJL results are compared  with the lattice data obtained at
fixed $m=0.05$, $\mu_I=0.4$ and $\lambda=0.005$ with
$N_t=4$\cite{Kogut:2002} in Fig.9(the PNJL results were calculated
with the same parameters). The critical $\beta_c$ at zero chemical
potential is 5.3198(9) for m=0.05 \cite{Kogut:2004}, which may
corresponds to $T_c^0=180\text{MeV}$. In lattice simulations with
fixed $N_t$, the temperature $T$ varies with lattice spacing $a$ and
$T=N_ta^{-1}$. Using 2-loop running of the coupling $g(a)$, the
lattice spacing $a$ can be expressed as
\begin{equation}
a=\Lambda_L^{-1}(\frac{6\beta_0}{\beta})^{-\frac{\beta_1}{2\beta_0^2}}\text{exp}(-\frac{\beta}{{12\beta_0}}),
\end{equation}
where $\beta=6/g^2,
\beta_0=\frac{1}{16\pi^2}(\frac{11N_c}{3}-\frac{2N_f}{3})$ and
$\beta_1=\frac{1}{(16\pi^2)^2}(\frac{34N_c^2}{3}-\frac{10N_cN_f}{3}-\frac{(N_c^2-1)N_f}{N_c})$.
The one dimensional constant parameter $\Lambda_L$ can be determined
by comparing the known critical $\beta_c$ with the corresponding
critical temperature $T_c^0$. Using this relation, we can transform
$Ts$ into $\beta$s and compare the PNJL results with the
corresponding lattice data. Note that in lattice simulations, every
quantity is made dimensionless by multiplying the appropriate power
of $a(\beta)$.
\par
Fig.9(a) shows that the dependence of Polyakov Loop on $\beta$ in
PNJL is well consistent with the lattice data. The peak of Polyakov
Loop susceptibility from PNJL is located at $\beta=5.27$, which
coincides with the lattice result depicted in Fig.14 in
Ref.\cite{Kogut:2002}. With 2-loop running of the coupling g, this
crossover transition point for deconfinement corresponds to
$T=169\text{MeV}$ and $\mu_I=270\text{MeV}$(note that $\mu_I$ also
changes with $\beta$) with $m=34\text{MeV}$. The variation of the
pion condensate with respect to $\beta$ is shown in Fig.9(b). Since
the original data for pion condensate in \cite{Kogut:2002} are all
dimensionless quantities scaled by a series of physical
units(corresponding to different lattice spacing $a(\beta)$), we
presented these data in physical units and then rescaled the
obtained results by the pion condensate at $\beta=5.2$. Both PNJL
and lattice data show that the pion condensate first increases with
$\beta$ and then decreases with increasing $\beta$. The crossover
phase transition for $I_3$ symmetry restoration in PNJL model takes
place at $\beta=5.3$ or ($T=175\text{MeV}$, $\mu_I=280\text{MeV}$)
with $m=35\text{MeV}$. In contrast with the critical value shown in
Fig.9(a), both the temperature and isospin chemical potential of
this point are shifted by less than $10\text{MeV}$. We also
performed calculations in PNJL for $m=0.05$ by keeping
$T_0=270\text{MeV}$ and adopting $T_c^0=222\text{MeV}$ as in
previous sections and found that both crossover transition points
located at the same point $\beta=5.3$. Therefore the coincidence of
the crossover transitions for deconfinement and $I_3$ symmetry
restoration obtained in lattice simulation
\cite{Kogut:2002,Kogut:2002} can be successfully interpreted by
including the Polyakov Loop dynamics in the NJL model.
\par
The $16^3\times4$ lattice simulations in \cite{Kogut:2004} with the
   set of same parameters as in Fig.9  strongly suggest
that there would be a first order phase transition for the $I_3$
symmetry restoration between $\beta=0.525$ and $\beta=5.3$ in the
$\lambda\rightarrow0$ limit. The authors of \cite{Kogut:2004}
claimed that this point is very close to the TCP.  This first order
phase transition was confirmed in our calculations within PNJL for
$\lambda=0$ and the transition point is located at $\beta=5.32$ or
($T=180\text{MeV}$, $\mu_I=288\text{MeV}$), which is also in the
vicinity of the TCP obtained in PNJL ($m=36\text{MeV}$).
\par
At fixed $\beta=5.0$ or $T=119\text{MeV}$, the pion condensate as a
function of $\mu_I$ for $m=0.05$ or m= 24\text{MeV} and different
$\lambda$ is plotted in Fig.10. The range of $\mu_I$ in this plot
corresponds to $0<\mu_I<239\text{MeV}$. The ellipses are the linear
extrapolation of the lattice data to $\lambda=0$ and the solid line
corresponds to the PNJL result obtained at $\lambda=0.0005$. Fig.10
indicates that the PNJL results are quite well consistent with
lattice data for $\lambda=0.01, 0.005$. This plot also shows that
the linear extrapolation to $\lambda=0$ in lattice simulation is
fairly good agreement with the PNJL data in the low and high $\mu_I$
region, while the deviations are explicit in the vicinity of the
true phase transition point. In contrast to the critical point
$\mu_I=0.25$(119\text{MeV}) obtained from the lattice linear
extrapolation, the critical isospin chemical potential in PNJL model
is $\mu_I=0.3$(143\text{MeV}). Both results sustain that the $I_3$
symmetry broken phase transition is second order for low temperature
(in contrast to $T_c^0$) and no decrease of pion condensate with
increase of $\mu_I$ is observed in the plotted range of $\mu_I$.
\begin{figure}[ht]
\centering \hspace{-.05\textwidth} \hspace{-.10\textwidth}
\scalebox{.45}
{\includegraphics*[width=1.17\textwidth]{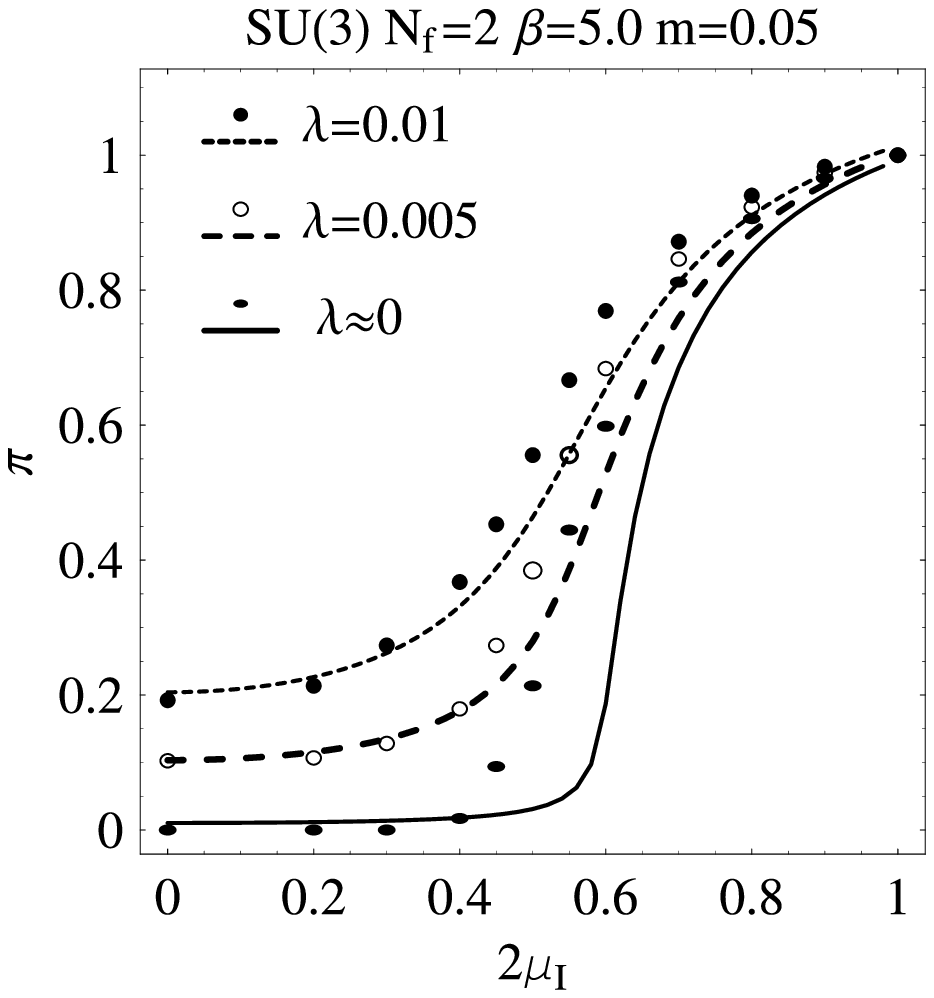}}\\
\centerline{}
\parbox{15cm}
{\caption{Scaled pion condensate as function of $\mu_I$ for
$\lambda=0.01$, $\lambda=0.005$ and $\lambda->0$ at fixed m=0.05 and
$\beta=5.0$. The lattice data are taken from \cite{Kogut:2002} and
we rescaled them by the lattice result for the pion condensate at
$2\mu_I=1$ and $\lambda=0.01$. The curves for the PNJL data are also
scaled by the  pion condensate obtained in PNJL at $2\mu_I=1$ and
$\lambda=0.01$. }} \label{fig10}
\end{figure}

\section{Conclusion and outlook}
\par
The PNJL model represents a minimal synthesis of the two basic
principles which govern QCD  at low temperatures: spontaneous chiral
symmetry breaking and confinement. In this work, we extended the two
flavor Polyokov-loop-extentded NJL model to the finite isospin
chemical potential case, which can be investigated by lattice
simulations. By including the pion condensate degrees of freedom,
the trace of Polyakov loop and its conjugate can still be extracted
out and the formalism which can simultaneously couple the fields of
the pion condensate, chiral condensate and Polyakov loop is
obtained.
\par
Within this framework, the dependence of the pion condensate, chiral
condensate and Polyakov loop on temperature and isospin chemical
potential are explored   and  the two flavor ($T,\mu_I$) phase
diagram is plotted at the mean field level. Our calculations
confirmed that there exists a tricritical point in the ($T,\mu_I$)
phase diagram. In the low $\mu_I$ region with pion condensate, the
crossover phase transitions for chiral symmetry and deconfinement
are still perfectly coincide and  the isospin symmetry restoration
phase transition for high isospin chemical potential region is
simultaneously accompanied with the crossover phase transition for
 deconfinement. We also directly compared the PNJL results with the corresponding
lattice data at finite temperatures. In general, a qualitative
agreement between these two methods is confirmed and in some cases
even a quantitative agreement is obtained .
 Therefore, our results provides further test on the validity of this enhanced
quasi-particle model. At the same time, the conclusions obtained
within PNJL model also give a way to test the reliability of the
lattice data for finite isospin density.
\par
In this paper, we only explored the ($T,\mu_I$) phase structure of
two flavor PNJL at zero baryon chemical potential. In our
forthcoming paper, we will investigate the Polyakov loop dynamics on
the phase structure, thermodynamic and meson properties at finite
temperature, isospin chemical potential and baryon chemical
potential. Further exploration on the phase diagram and meson
properties of  three flavor PNJL model by including the strangeness
chemical potential is also underway.

\vspace{5pt}
\noindent{\textbf{\Large{Acknowledgements}}}\vspace{5pt}\\
We are highly grateful to Dr. D.K. Sinclair for his helpful
discussions on the detailed technique of lattice QCD simulations.
 This work was supported by National Natural Science
Foundation of China under contract 10425521, 10575004 and 10675007,
the key Grant Project of Chinese Ministry of Education (CMOE) under
contract No.305001 and the Research Fund for the Doctoral Program of
Higher Education of China under Grant No.20040001010. One of the
author (LYX) thanks also the support of the Foundation for
University Key Teacher by the CMOE.

\end{document}